\documentclass[useAMS,articles,fleqn] {mn2e}
\usepackage{epsfig}
\usepackage{amsmath}
\usepackage{graphicx}
\usepackage{amssymb}

\title[IMF of young stellar clusters: bayesian analysis]{Testing the universality of the IMF with Bayesian statistics: young  clusters}

\author[Sami Dib]{Sami Dib$^{1,2}$\thanks{E-mail: sdib@nbi.dk}\\ 
$^{1}$Niels Bohr International Academy, Niels Bohr Institute, Blegdamsvej 17, DK-2100, Copenhagen, Denmark\\ 
$^{2}$Centre for Star and Planet Formation, University of Copenhagen, {\O}ster Voldgade 5-7., DK-1350, Copenhagen, Denmark\\ 
}

\begin{document}
\maketitle

\date{Accepted XXX. Received XXX}

\pagerange{\pageref{firstpage}--\pageref{lastpage}}
\pubyear{2009}
\label{firstpage}

\begin{abstract} 

The universality of the stellar initial mass function (IMF) is tested using Bayesian statistics with a sample of eight young Galactic stellar clusters (IC 348, ONC, NGC 2024, NGC 6611, NGC 2264, $\rho$ Ophiuchi, Chameleon I, and Taurus). We infer the posterior probability distribution function (pPDF) of the IMF parameters when the likelihood function is described by a tapered power law function, a lognormal distribution at low masses coupled to a power law at higher masses, and a multi-component power law function. The inter-cluster comparison of the pPDFs of the IMF parameters for each likelihood function shows that these distributions do not overlap within the $1\sigma$ uncertainty level. Furthermore, the most probable values of the IMF parameters for most of the clusters deviate substantially from their values for the Galactic field stellar IMF. We also quantify the effects of taking into account the completeness correction as well as the uncertainties on the measured masses. The inclusion of the former affects the inferred pPDFs of the slope of the IMF at the low mass end  while considering the latter affects the pPDFs of the slope of the IMF in the intermediate- to high mass regime. As variations are observed in all of the IMF parameters at once and for each of the considered likelihood functions, even for completeness corrected samples, we argue that the observed variations are real and significant, at least for the sample of eight clusters considered in this work. The results presented here clearly show that the IMF is not universal. 

\end{abstract} 


\section{INTRODUCTION}\label{intro}

Probing the universality of the stellar initial mass function (IMF, i.e., the distribution of the masses of stars at their birth), is one of the most challenging issues in modern astrophysics. The shape of the IMF and its potential dependence on the environment are crucial to almost all branches of modern astrophysics ranging from planetary science, the evolution of stellar clusters, the dynamics and chemical enrichment of the interstellar medium, and galactic evolution. One of the first attempts to determine the shape of the IMF was made by Salpeter (1955) who derived the mass function of nearby Galactic field stars of masses $0.4 < M_{\star}/M_{\odot} < 10$ and found that it is well described by a power law $dN/d{\rm log}M=M^{-\Gamma}$, where $N$ is the number of stars between ${\rm log}M$ and ${\rm log}M+d{\rm log}M$ and with $\Gamma \approx 1.35$. Miller \& Scalo (1979) constructed the mass function of stars in the solar neighborhood and fitted it with a log-normal function with a steeper-than Salpeter slope in the high mass regime (i.e., $\Gamma \approx 1.7$). They also extended the IMF to the low mass regime and found the numbers of solar and sub-solar mass stars to fall below the expectations of the extrapolated Salpeter mass function. Scalo (1986) combined Galactic field stars with OB associations and obtained a mass function which is also steeper than the Salpeter mass function (i.e, $\Gamma \approx 1.7$) for $M_{\star} \gtrsim 2$ M$_{\odot}$ (see also Rana 1987). A three component power-law fit of the Galactic field mass function has been proposed by Kroupa et al. (1993), Scalo (1998) and Kroupa (2001,2002) whereas Chabrier (2003) suggested that the Galactic field stellar mass function can be fitted by a lognormal distribution for masses $\lesssim 1$ M$_{\star}$ and by the Salpeter mass function for $> 1$ M$_{\star}$. Bochanski et al. (2010) constructed the mass function of 15 million low-mass stars ($0.1 < M /{\rm M}_{\odot} < 0.8$) and suggested it can be well represented by a log-normal function which peaks at $\approx 0.8$ M$_{\odot}$. Hollenbach et al. (2005) and more recently Parravano et al. (2011) proposed that the Galactic field mass function is well described by a single functional form across the entire range of stellar masses which is a Rosin-Rammler function tapered by a Salpeter-like power-law. Other authors have proposed other functional forms such as the product of a power law and an exponential function (Larson 1998), and an order-3 Logistic function (Maschberger 2013).      

\begin{figure*}
\begin{center}
\includegraphics[height=13cm, width=15cm]{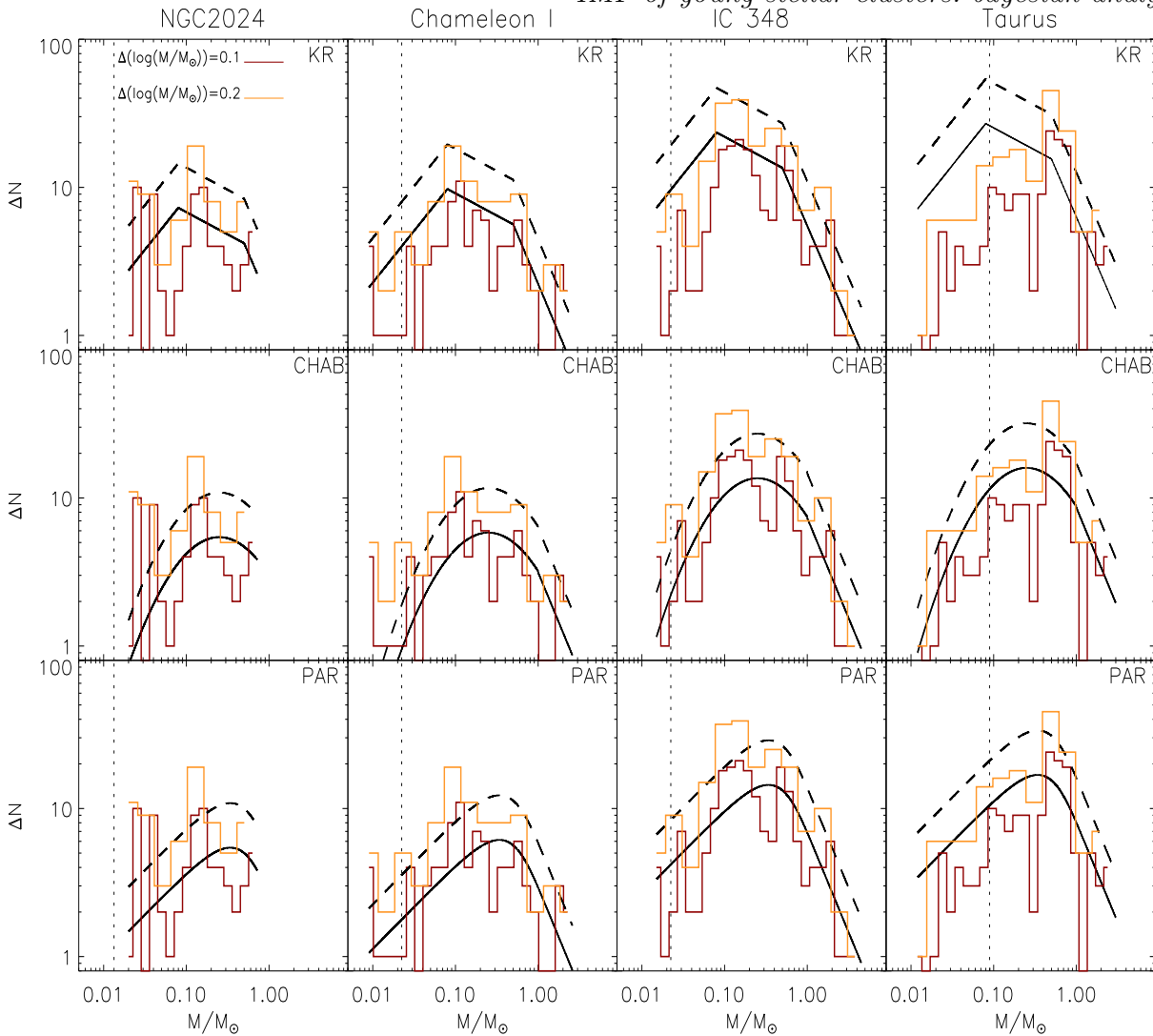}
\end{center}
\vspace{0.5cm}
\caption{The number of stars $\Delta N$ per logarithmic bin size $\Delta({\rm log}(M/{\rm M}_{\odot}))$ of the young stellar clusters NGC 2024, Chameleon I, IC 348, and Taurus. Stellar masses are binned with logarithmic bin sizes of $\Delta({\rm log}(M/{\rm M}_{\odot}))=0.1$ M$_{\odot}$ (red lines) and $\Delta({\rm log}(M/{\rm M}_{\odot}))=0.2$ (yellow line). The full and dashed blacks lines represent the Kroupa (2001) mass function (first row), the Chabrier (2005) mass function (middle row), and the Parravano et al. (2011) mass function (bottom row) with their fiducial Galactic IMF parameter. All functions are normalized to the total mass contained in each cluster (see text for more details).}
\label{fig1}
\end{figure*}

\begin{figure*}
\begin{center}
\includegraphics[height=13cm, width=15cm]{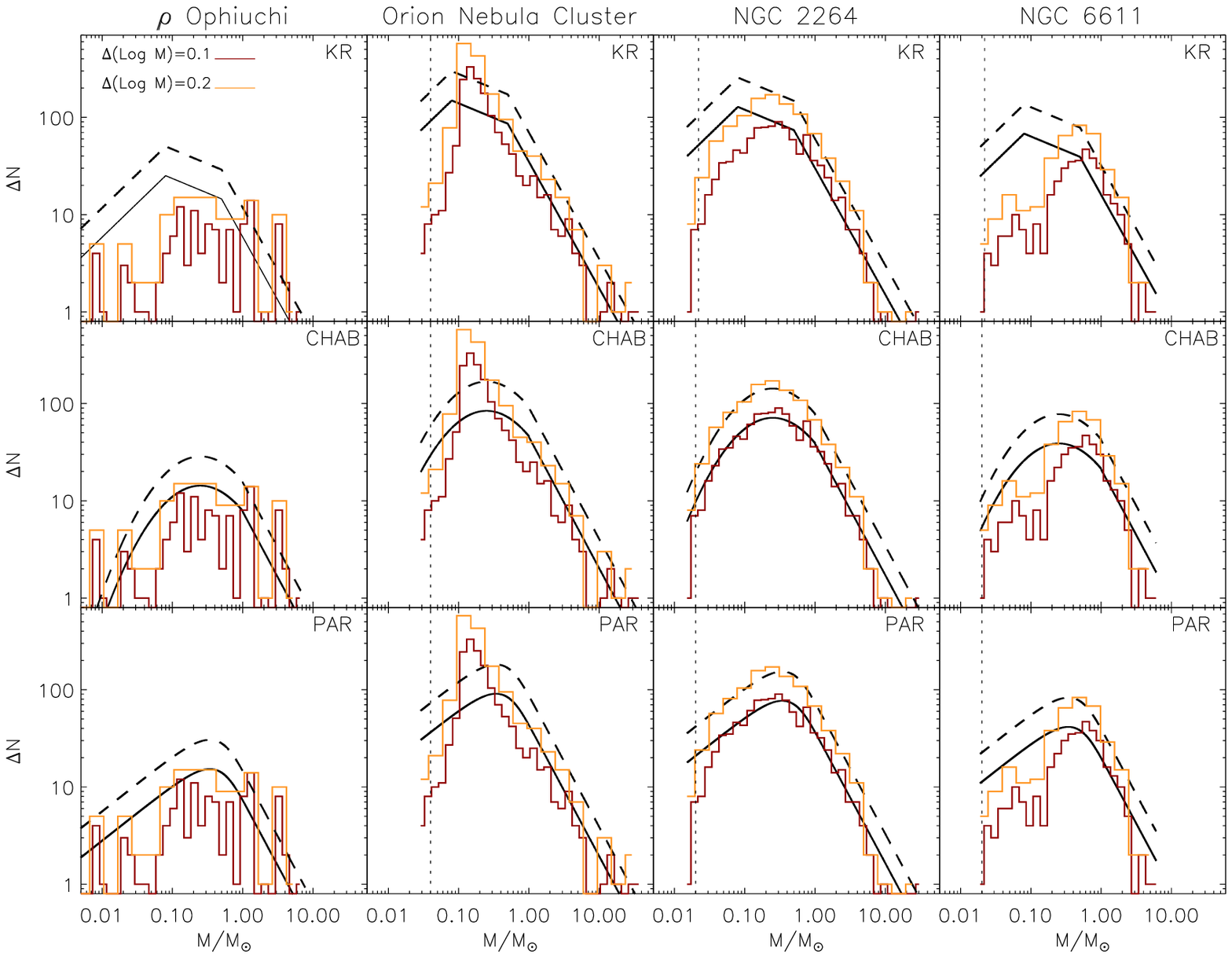}
\end{center}
\vspace{0.5cm}\caption{Same as Fig.~\ref{fig1} but for the young clusters $\rho$ Ophiucus, Orion Nebula Cluster, NGC 2264, and NGC 6611.}
\label{fig2}
\end{figure*}

The mass function of stars that can be counted in the field of the Galaxy includes stars with a wide spread in age, that have various metallicities, and that were formed in regions of potentially differing properties. In contrast, stars in clusters have roughly the same age, metallicity, and are located at the same distance. Thus, one can presume that their observed present day mass functions (PDMFs\footnote{In the remainder of the paper, we will refer to the PDMF, especially to that of young clusters as being the IMF. However, it should always be kept in mind that we are dealing here with PDMFs. }) are a fair representation of their IMFs especially in the case of young clusters in which the effects of stellar and dynamical evolution are minimal. The IMF have been constructed for a large number of clusters in the Galaxy and in the Magellanic Clouds including open clusters (Massey et al. 1995; Massey \& Hunter 1998; Massey 1998; Sanner \& Geffert 2001; Slesnick et al. 2002; Piskunov et al. 2004; Moraux et al. 2004; Gouliermis et al. 2005; Pandey et al. 2007; Sharma et al. 2008; Silva-Villa et al. 2008; Bouvier et al. 2008; Liu et al. 2009; Boudreault et al 2010; Baker et al. 2010; Casewell et al. 2011; Wang et al. 2011; Lodieu et al., 2011), globular clusters (Paresce \& De Marchi 2000; De Marchi et al 2010; Paust et al. 2010), and young clusters and stellar associations (Massey et al. 1989; Massey \& Thompson 1991; Massey \& Johnson 1993; Parker et al. 1992; Hillenbrand et al. 1993; Oey 1996; Hillenbrand 1997; Herbig 1998; Sirianni et al. 2000,2002; B\'{e}jar et al. 2001; Muench et al. 2002; Preibisch et al. 2002; Liu et al. 2003; Luhman 2000,2004a,b,2007; Luhman et al. 1999;2003a,2003b,2009; Selman \& Melnick 2005; Prisinzano et al. 2005; Oliveira et al. 2005,2008; Leistra et al. 2005,2006; Levine et al. 2006; Massi et al. 2006; Schmalzl et al. 2008; Sung et al. 2008; Andersen et al. 2008; Harayama et al. 2008; Espinoza et al. 2009; Scholz et al. 2009; Sung \& Bessel 2010; Campbell et al. 2010; Salas \& Cruz-Gonz\'{a}lez 2010; Ojha et al. 2010; Delgado et al. 2011; Bayo et al. 2011; Gennaro et al. 2011; Da Rio et al. 2012a; Alves de Oliveira et al. 2012, Tripathi et al. 2014, Mallick et al. 2014). The derivation of the shape of the PDMF in these studies relied on constructing the histogram of stellar masses (or of their logarithmic values) in bins of equal sizes and then fitting it with one or several functional forms. A minimization of the chi-square of the fit then allows for the derivation of the fit parameters and their associated uncertainties. In the intermediate- to high mass regimes, the values of the slope derived using this approach vary in the range between $0.7$ and $2$ when stellar masses are binned logarithmically. It is often claimed in the literature that, within the $1\sigma$ uncertainty, the derived values are broadly consistent with the Salpeter slope. However, this is far from being clear. An inspection of the derived values of the slope in the this mass regime for several clusters for which identical data reduction algorithms and theoretical evolutionary tracks have been used to derive the stellar masses does suggest that the slopes of the IMF for these clusters do not overlap within the $1\sigma$ uncertainty level (e.g., Massey 2011; Sharma et al. 2008; Lata et al. 2010; Tripathi et al. 2014).

Deriving the slopes/shape of the IMF in different mass regimes using binned data is not without problems. One major issue is the dependence of the derived slopes on the size of the bin, particularly when dealing with low number statistics (Ma\'{i}z Apell\'{a}niz \& \'{U}beda 2005). Ma\'{i}z Apell\'{a}niz \& \'{U}beda (2005) also showed that the use of uniform size bins in mass or the logarithm of the mass can bias the determination of the slope of the IMF when using a $\chi^{2}$ minimization of the binned data. The origin of the bias lies in the correlation between the number of stars per bin and the weights assigned to each bin. They showed that these biases can be reduced by using bins of variable sizes which contain equal numbers of stars. However, a potential danger of this method is to smooth features in the IMF (e.g., bumps of physical origin) by forcing stars to be distributed in bins containing equal numbers of stars. Another important limitation of fitting binned mass functions is the undesired effect of the subjective choice of break points (i.e., the points where the mass functions turns from one power law/functional form to another). In this work, we use a Bayesian statistics approach to infer the parameters that characterize the shape of the PDMF/IMF for a number of young stellar clusters when the underlying IMF is assumed to be described by the tapered power law function, a lognormal function at low masses coupled to a power law function at higher masses, and a three-component power law function. Allen et al. (2005) applied a Bayesian analysis in order to infer the slope of the low mass end of the Galactic field IMF. However, the observational data did not allow them to distinguish between a single power law, a two-segment power law, or a lognormal function. Olmi et al. (2014) used a similar approach to infer and compare the parameters describing the shape of the mass function of submillimeter clumps in two {\it Herschel infrared GALactic Plane Survey} fields. Weisz et al. (2013) used mock data to quantify the uncertainties associated with the inference of the high mass slope of the IMF by Bayesian methods, as a function of the number of stars in the cluster and the range of stellar masses involved. They used this to argue that the uncertainties on the slopes of the IMF at the high mass end derived using standard least square fitting methods tend to be generally underestimated. In \S.~\ref{models}, we write down the three IMF models we intend to compare the observational data with. In \S.~\ref{observations}, we briefly present and discuss the basic features of the observational data we employ for a number of young clusters. The Bayesian inference method is presented in \S.~\ref{bayesian}. The inter-cluster comparison for each of the considered IMF models in presented in \S.~\ref{intercluster} and an inter-model comparison for each cluster is presented in \S.~\ref{intermodel}. In \S.~\ref{completeness} the effects of completeness and uncertainties on stellar masses are presented. In \S. ~\ref{discussion} we discuss some of the issues related to variations in the IMF, and in \S.~\ref{conclusions}, we conclude.

\begin{figure}
\begin{center}
\includegraphics[width=\columnwidth]{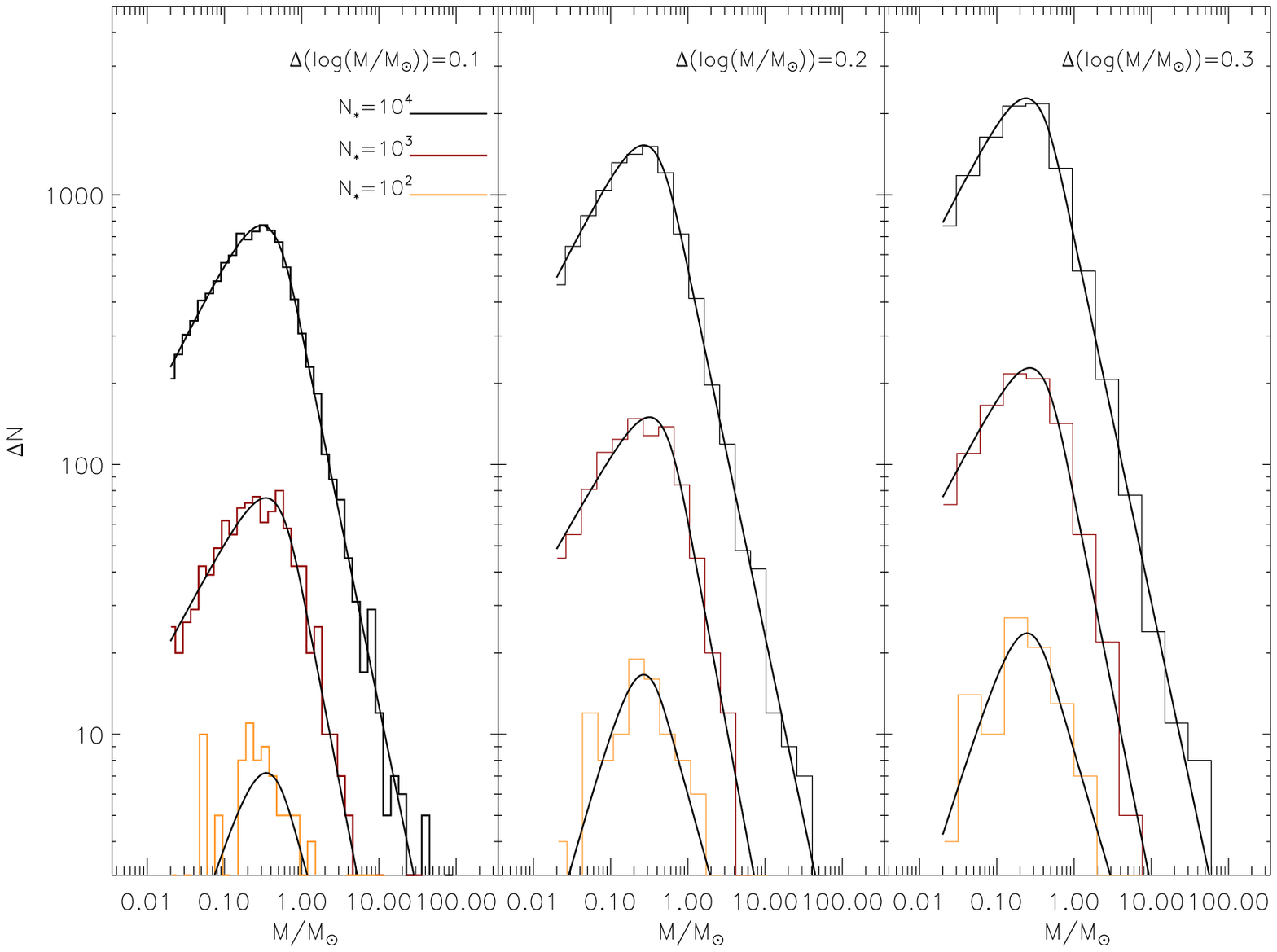}
\end{center}
\vspace{0.3cm}
\caption{Synthetic IMFs constructed using the Parravano et al. (2011) IMF with the Galactic field values of the parameters ($\gamma_{P}=0.57$, $M_{P}=0.42$M$_{\odot}$, $\Gamma_{P}=1.35$) for three clusters with a number of stars $N_{\star}=10^{2}$, $10^{3}$, and $10^{4}$ and for masses in the range [0.02-150] M$_{\odot}$. The drawn masses for each cluster are binned using logarithmic bins with sizes of ${\rm log}(M/{\rm M}_{\odot})$=0.1(left panel), ${\rm log}(M/{\rm M}_{\odot})=0.2$ (middle panel), and ${\rm log}(M/{\rm M}_{\odot})=0.3$ (right panel). Over-plotted to each binned mass function (full black line) is the Parravano et al. IMF with the fiducial Galactic field values of the IMF parameters.}
\label{fig3}
\end{figure}

\begin{figure}
\begin{center}
\includegraphics[width=0.9\columnwidth]{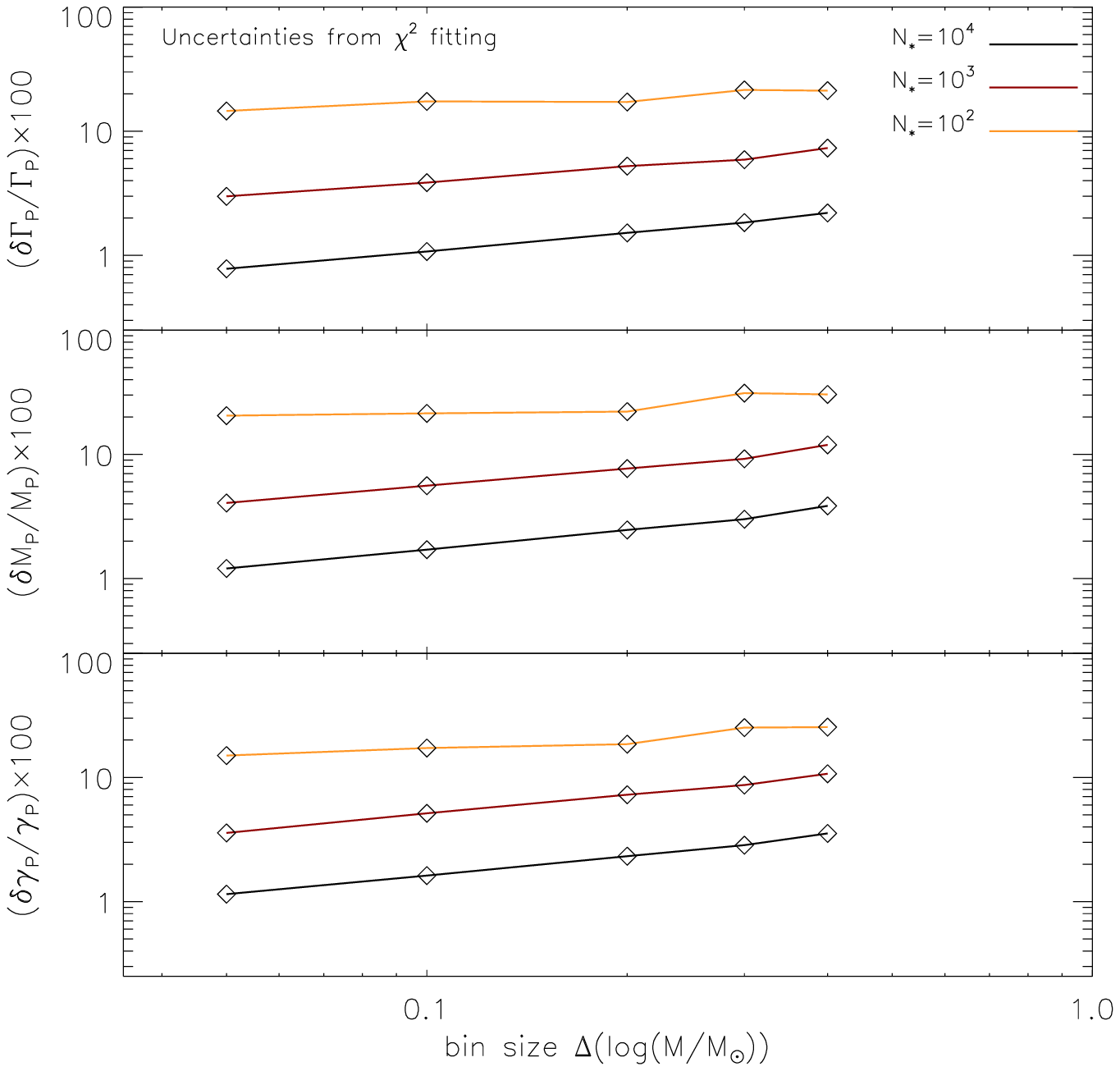}
\end{center}
\vspace{0.5cm}
\caption{Dependence of the 1$\sigma$ uncertainty of the IMF parameters on the bin-size for three clusters with a number of stars $N_{\star}=10^{2}$, $10^{3}$, and $10^{4}$. The stellar masses are  from the Parravano et al. (2011) field IMF (i.e., the ones shown in Fig.~\ref{fig3}) and the values and uncertainties of the IMF parameters are derived using a standard $\chi^{2}$ minimisation technique. The figure shows clearly the dependence of the uncertainties on the adopted bin size.}
\label{fig4}
\end{figure}

\section{IMF models}\label{models}

Some of the most commonly used functional forms in the literature to describe the IMF are a multi-component power law function (Kroupa 2001,2002), a lognormal distribution for masses $\le 1$ M$_{\odot}$ complemented with a Salpeter slope for masses $\ge 1$ M$_{\odot}$ (Chabrier 2003,2005), and a tapered power-law function (De Marchi et al. 2010, Parravano et al. 2011). The sets of parameters that define the Galactic field IMF in each of the models (below) refer to a system IMF, that is, when the census of stars is uncorrected for unresolved binary stars. Parravano et al. (2011) applied the tapered power-law IMF (hereafter referred to as PAR) to the Galactic field observations of Reid et al (2002). They found that the system IMF can be described by: 

\begin{equation}
\psi_{P}\left({\rm log}M \right)=A_{P} M^{-\Gamma_{P}}  \{1-{\rm exp}[-(M/M_{P})^{\gamma_{P}+\Gamma_{P}}]\}, 
\label{eq1}
\end{equation}

\noindent where $A_{P}$ is a normalisation constant and $\Gamma_{P}=1.35$, $M_{P}=0.42$ M$_{\odot}$, and $\gamma_{P}=0.57$. Chabrier (2005) combined data in the V-band from Reid et al. (2002;2004) and K-band data from Henry \& McCarthy (1990) to construct the Galactic field IMF which he parametrized as (referred to in the text as CHAB): 

\begin{eqnarray}
\begin{array}{llll}
 \psi_{C}\left({\rm log}M \right)&=& A_{C}\times0.076\times{\rm exp}\left\{ -\frac{({\rm log}M - {\rm log}M_{c})^{2}}{2\sigma_{c}^{2}} \right \}  &, M \leq  M_{br} \\
                                                     &=& A_{C}\times0.041\times M^{-\Gamma_{c}}                                                                                                           &, M \geq  M_{br},
\end{array}                                               
\label{eq2}
\end{eqnarray}

\noindent where $A_{C}$  is a normalization constant, $M_{C}=0.2$ M$_{\odot}$, $\sigma_{C}=0.55$ M$_{\odot}$, $\Gamma_{C}=1.35$, and $M_{br}=1$ M$_{\odot}$. Finally the multi-component power law IMF s described by (Kroupa 2002; Weidner \& Kroupa 2004, referred to in the text as KR): 

\begin{eqnarray}
\begin{array}{l} 
 \psi_{K}\left({\rm log}M \right)=A_{K} 
 \\
 \end{array}
 \left\{
 \begin{array}{l}
 M_{K1}\left(\frac{M} {M_{K1}}\right)^{-\Gamma_{K1}}, M \leq M_{K1} \\ 
M_{K1}\left(\frac{M} {M_{K1}}\right)^{-\Gamma_{K2}}, M_{K1} \leq M  \leq M_{K2}\\
M_{K2}\left(\frac{M_{K2}}{M_{K1}}\right)^{-(1+\Gamma_{K2})}\left(\frac{M}{M_{K2}}\right)^{-\Gamma_{K3}}, M \geq M_{K2}  \\
\end{array}
\right.  
\label{eq3}
\end{eqnarray}

\noindent with $M_{K1}=0.08$ M$_{\odot}$, $M_{K2}$=0.5 M$_{\odot}$, $\Gamma_{K1}=-0.7$, $\Gamma_{K2}=0.3$, and $\Gamma_{K3}=1.35$.

\section{OBSERVATIONAL SAMPLE OF CLUSTERS}\label{observations} 

\begin{table*}
\begin{center}
\caption{Name of the clusters followed by the number of considered stars ($N_{\star}$), the total mass of the stars considered in the cluster $(M_{cl})$, the minimum and maximum stellar masses found in the cluster ($M_{min}$ and $M_{max}$, respectively), The completeness limit $M_{comp}$, and the estimated age of the cluster.}

\begin{tabular}{lcccccl}
\hline
\hline
Cluster     & $N_{\star}$   &  $M_{cl}$                            &  $M_{min}-M_{max}$  & $M_{comp}$       &  age\footnotemark[1]                    & Reference \\ 
                  &                        &   [M$_{\odot}$]                    &  [M$_\odot$]                   &   [M$_{\odot}$]   & [Myr]                   &                       \\   
\hline
Taurus      &      159           &   87.01                                &    0.012-3                       &       $0.02$            &     $\approx 1$   &   Luhman et al. (2009)      \\
Cha I         &         85          &   30.43                                 &   0.009-2.6                    &       $0.03$            &    $\approx 3-4$       &          Luhman (2007)               \\
IC 348       &      192           &   82.22                                &   0.015-4.5                     &       $0.03$            &    $\approx 2$     &           Luhman et al. (2003a)  \\  
ONC          &     1519          &   693.5                                &   0.029-45.7                  &        ${\rm individual}$          &     $0.3-1$  &           Da Rio  et al. (2012a)    \\
$\rho$ Ophiuchi  &   114  &    96.49                               &   0.003-7.72                  &        $0.003$         &       $0.3-1$ &     Alves de Oliveira et al. (2012) \\
NGC 6611 &            355    &    250.3                              &  0.019-6.04                   &        $0.022$          &   $\approx 2-3$       &           Oliveira et al. (2009)     \\
NGC 2264 &            990    &    574.8                              &  0.015-33.8                   &        $0.25$\footnotemark[2]        &    $0.3-2$      &           Sung \& Bessel (2010)  \\
NGC 2024 &              69    &    13.9                                 &  0.02-0.72                     &        $0.04$            &     $\approx 0.5$      &          Levine et al. (2006)        \\ 

\hline 
\end{tabular}

\noindent \footnotemark[1]{All clusters studied here contain stellar populations that have an age spread of variable value. The ages reported here refer to the peak of the distribution of stellar ages.}\\
\noindent \footnotemark[2]{This high value of $M_{comp}$ is due to a cluster membership criterion that is based on the existence of X-ray emission. }

\label{tab1}
\end{center}
\end{table*}

\begin{figure*}
\centering
\begin{tabular}{cc}
\epsfig{file=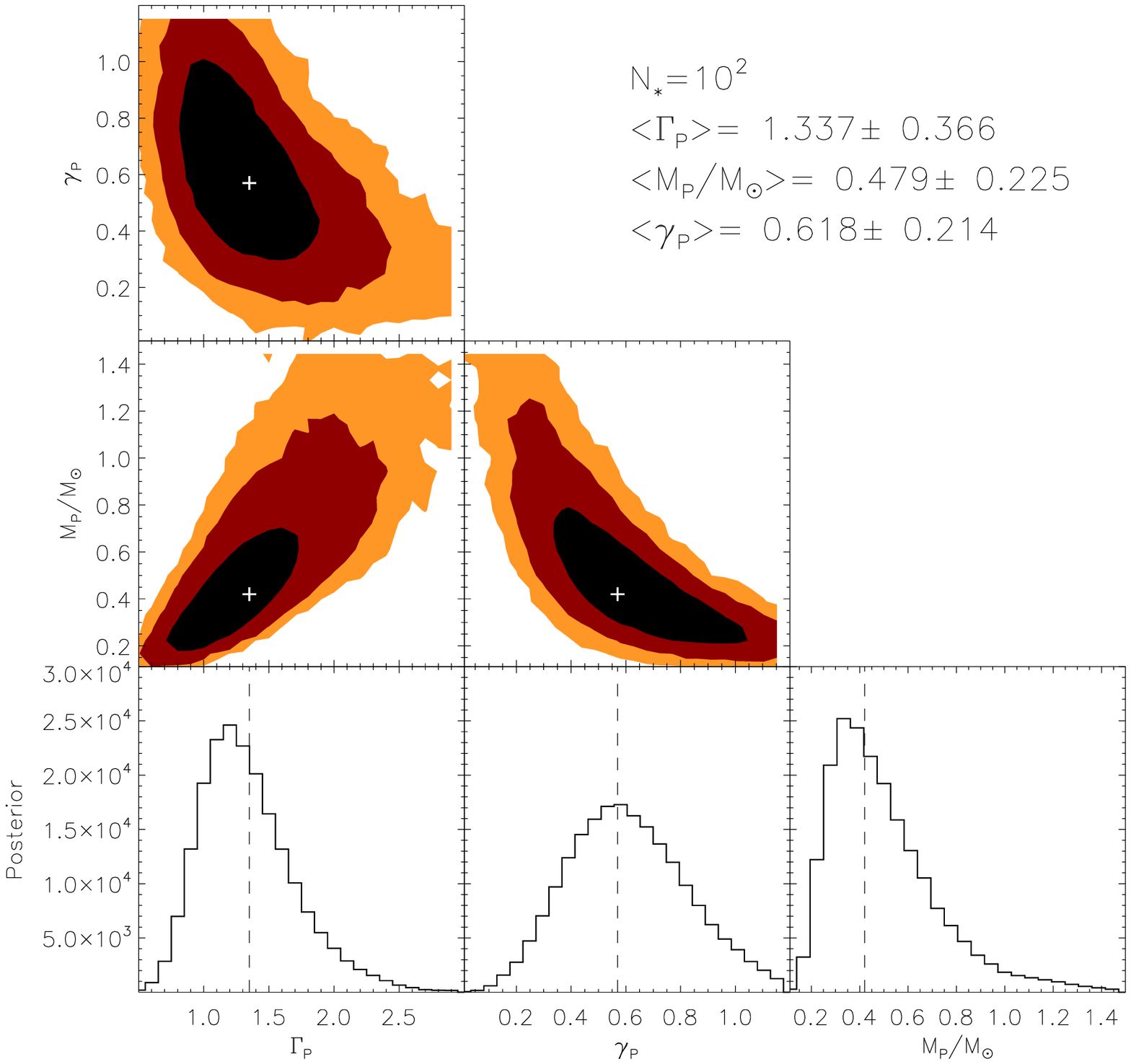,width=0.4\textwidth} &
\vspace{1cm}
\hspace{0.5cm} 
\epsfig{file=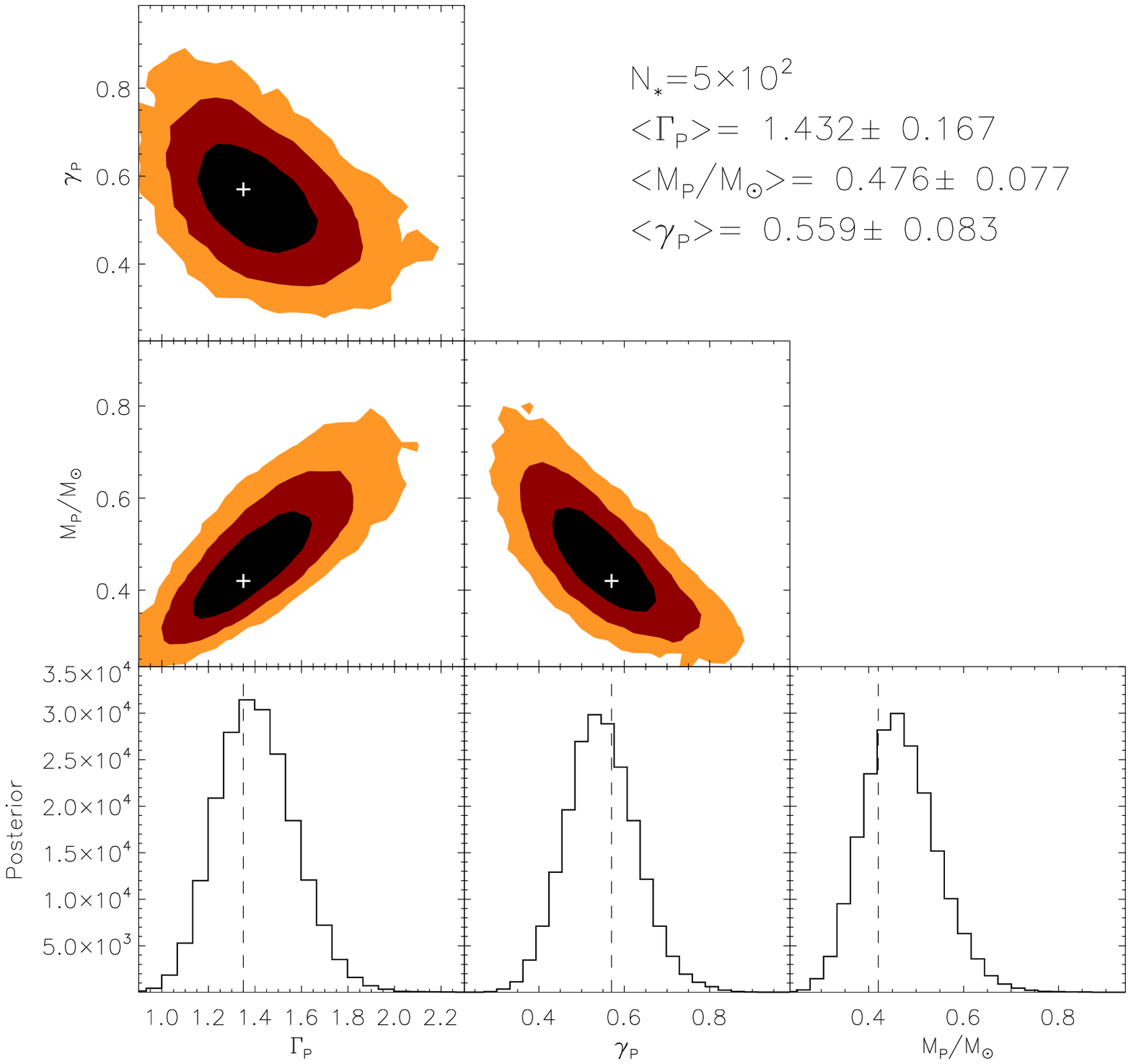,width=0.4\textwidth} \\
\epsfig{file=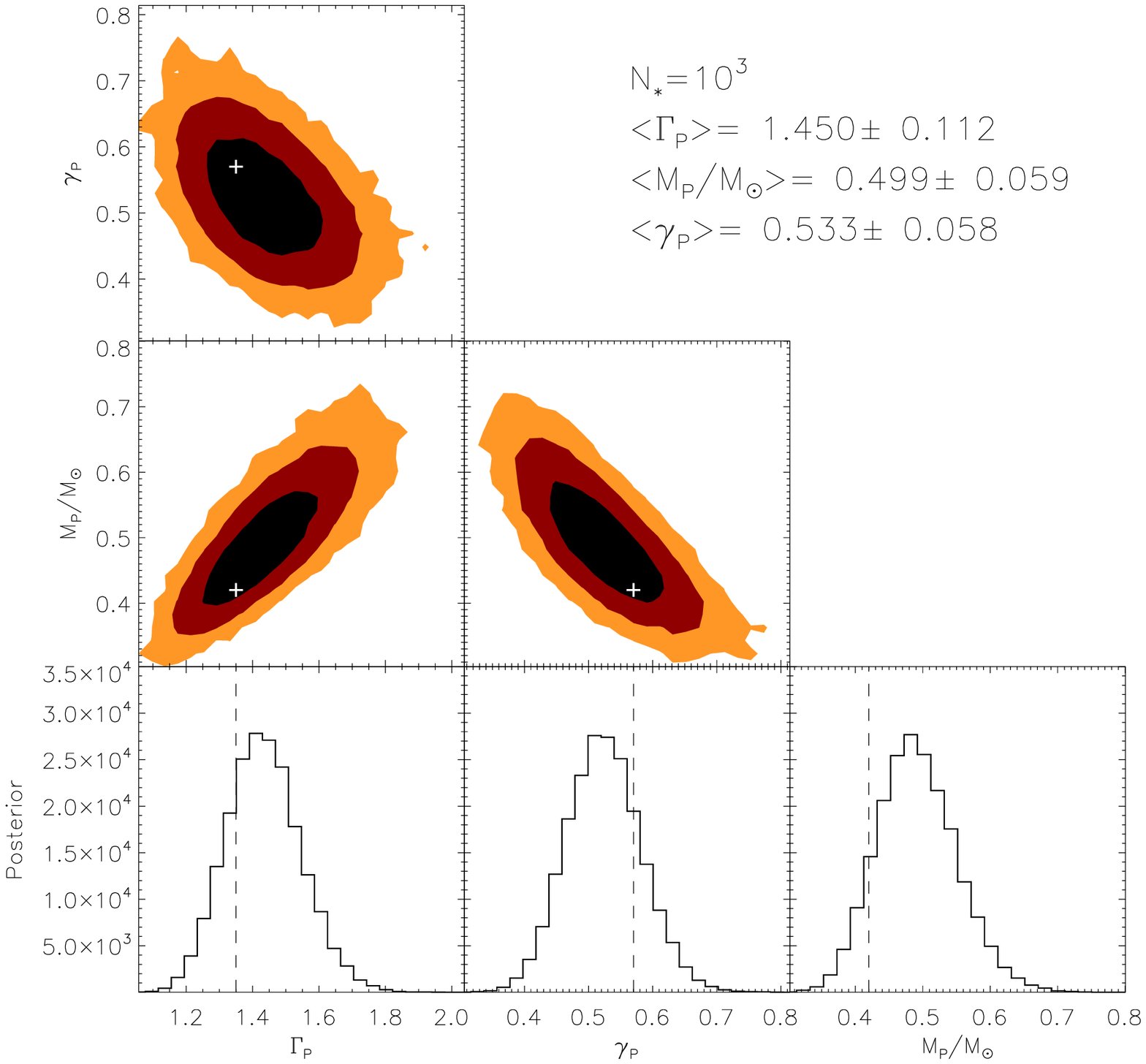,width=0.4\textwidth} &
\vspace{1cm}
\hspace{0.5cm}
\epsfig{file=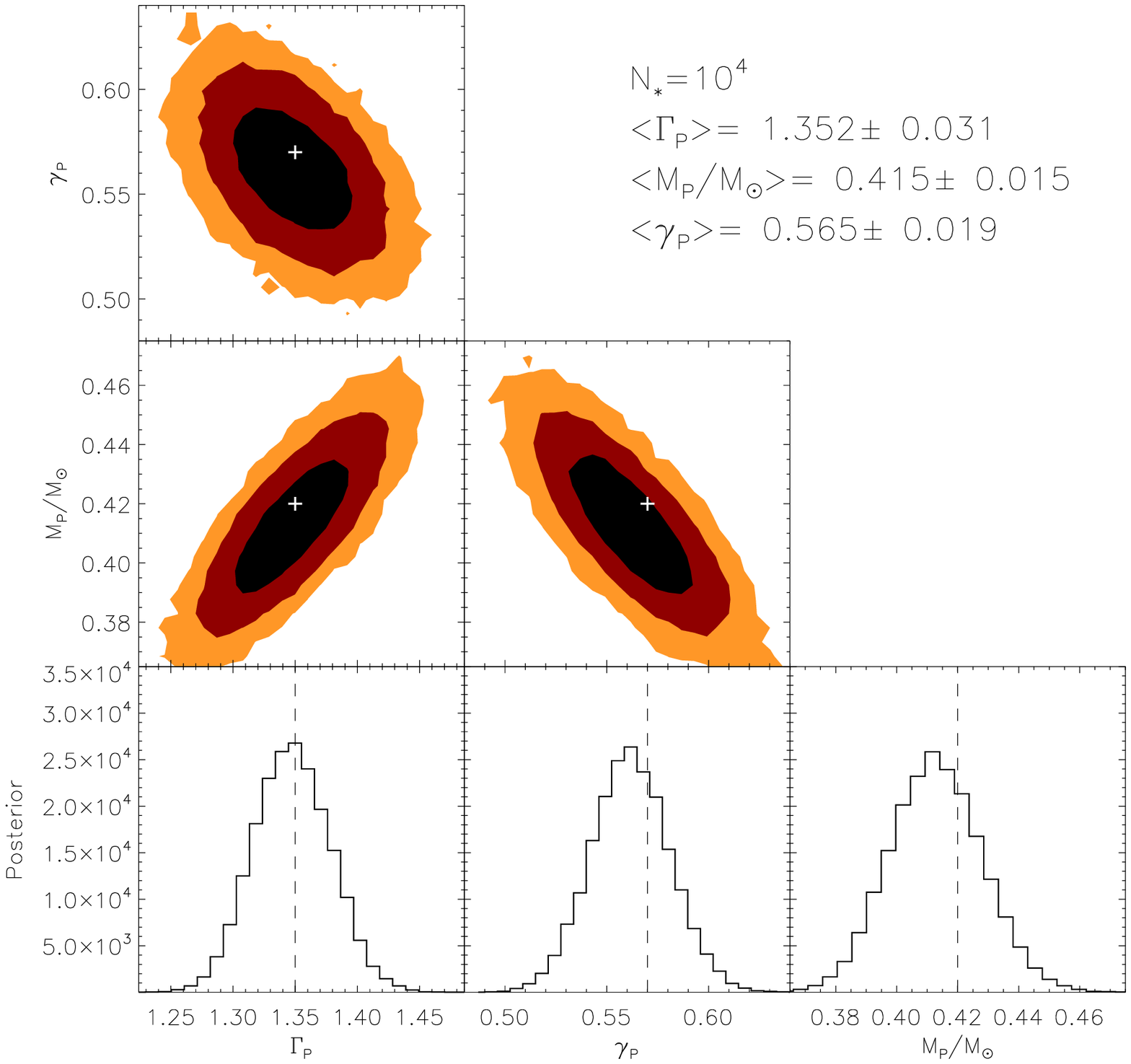,width=0.4\textwidth}
\end{tabular}
\caption{Posterior probability density functions (pPDF, bottom row) and 2D density probability functions of the IMF parameters for four synthetic clusters with $N_{\star}=10^{2}$ (top left), $N_{\star}=5\times10^{2}$ (top right), $N_{\star}=10^{3}$ (bottom left), and $N_{\star}=10^{4}$ (bottom right) when the likelihood function is described by the tapered power law function (Eq.~\ref{eq1}). The ensemble of stars for which the IMF parameters have been inferred have been constructed using the tapered power law function with values of the parameters that are those derived for the Galactic field (Parravano et al. 2012) ($\gamma_{P}=0.57$, $M_{P}=0.42$ M$_{\odot}$, and $\Gamma_{P}=1.35$). The 1D pPDFs and the 2D density distributions of the parameters are binned in 25 bins. The latter are shown at the $1\sigma$ (black contours), $2\sigma$ (red contours), and $3\sigma$ (yellow contours) confidence levels. Note that the axes are different in each panel. The Galactic field values of the parameters are shown with the white crosses in the 2D figures and with the black dashed lines in the 1D figures. The figure shows the ability of the Bayesian method with an MCMC approach in recovering the injected values of the parameters. The mean values of the parameters displayed in the upper right corner of each panel are shown for the sake of comparison only and are not necessarily representative of the model with the highest posterior probability.}
\label{fig5}
\end{figure*}

\begin{figure*}
\begin{center}
\includegraphics[width=0.9\textwidth,height=15cm]{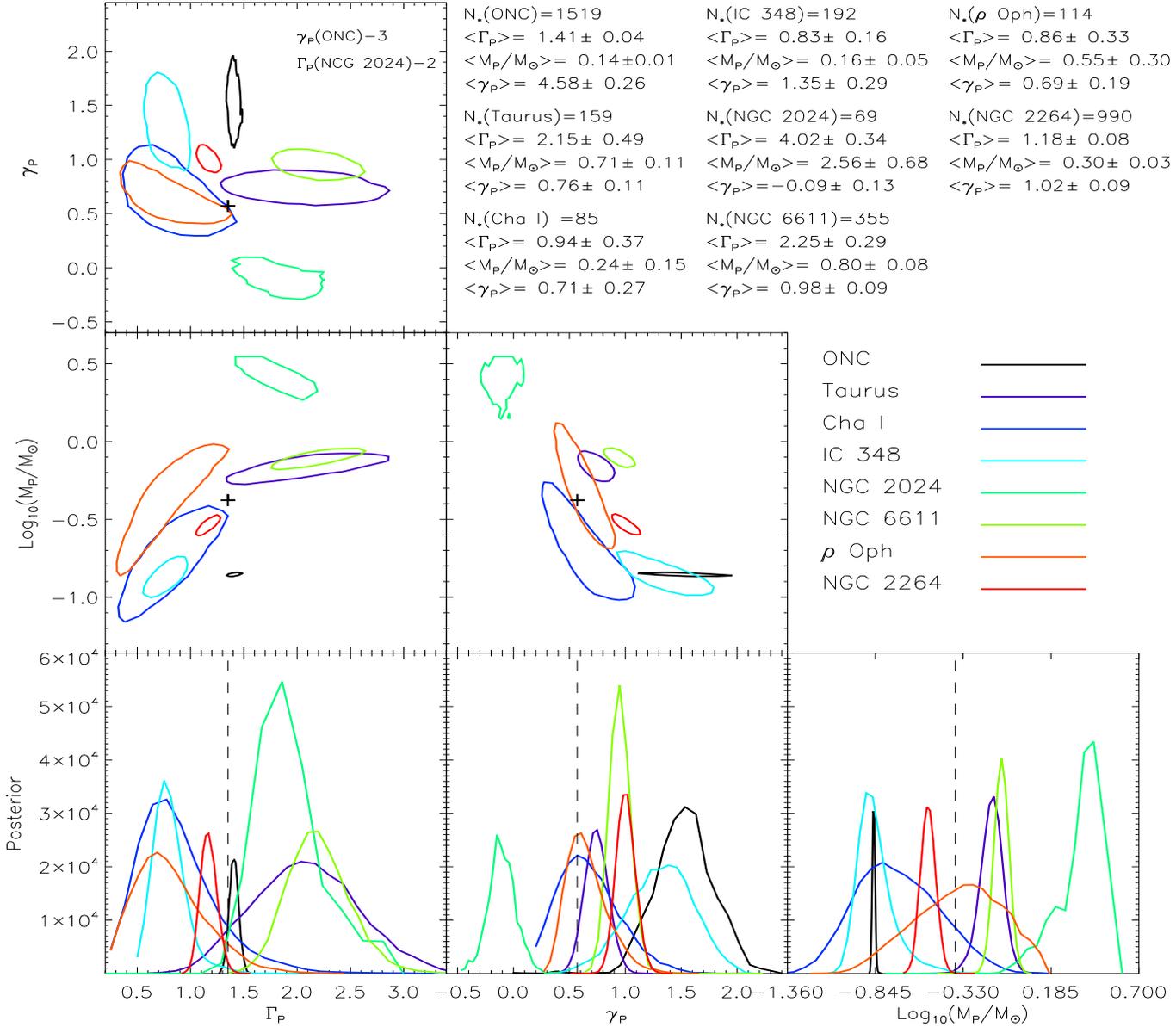}
\end{center}
\vspace{0.5cm}
\caption{Posterior probability density functions (pPDFs, bottom row) and 2D density probability functions of the parameters that describe the IMF for the eight young stellar clusters considered in this work when the likelihood function is given by the tapered power law function (Eq.~\ref{eq1}). The three parameters are: the slope at the low mass end $\gamma_{P}$, the characteristic mass $M_{P}$, and the slope at the high mass end $\Gamma_{P}$. The posterior distributions in the $(\Gamma_{P}-\gamma_{P})$, ($\Gamma_{P}-M_{P}$), and ($\gamma_{P}-M_{P}$) spaces are shown at the $1\sigma$ level. The Galactic field values of the parameters derived by Parravano et al. (2012) for the Galactic field IMF are shown with the black crosses in the 2D figures and with the black dashed lines in the 1D figures. The mean values of the parameters displayed in the upper right corner of the figure are shown for the sake of comparison only and are not necessarily representative of the model with the highest posterior probability.}
\label{fig6}
\end{figure*}

In this work, we infer the parameters that describe the IMF of a number of young stellar clusters  with relatively well resolved stellar populations\footnote{Still mostly uncorrected for binarity, so their IMF can be considered a system IMF.} and in which a turnover in the IMF at low masses has already been established. This allows us to use the entire mass range when deriving the parameters that characterize their IMFs and when comparing them to the Galactic field IMF models. The clusters considered in this work have been primarily selected based on the availability of their data (i.e., individual stellar masses). In order to minimize the effects of systematic differences in the derived masses of individual stars, we have given preferences to the mass derivation that employs, as much as possible, the same theoretical stellar evolutionary tracks. In this study, we consider eight clusters whose basic properties are summarized in Tab.~\ref{tab1}. These are:    

\begin{figure*}
\begin{center}
\includegraphics[width=0.9\textwidth]{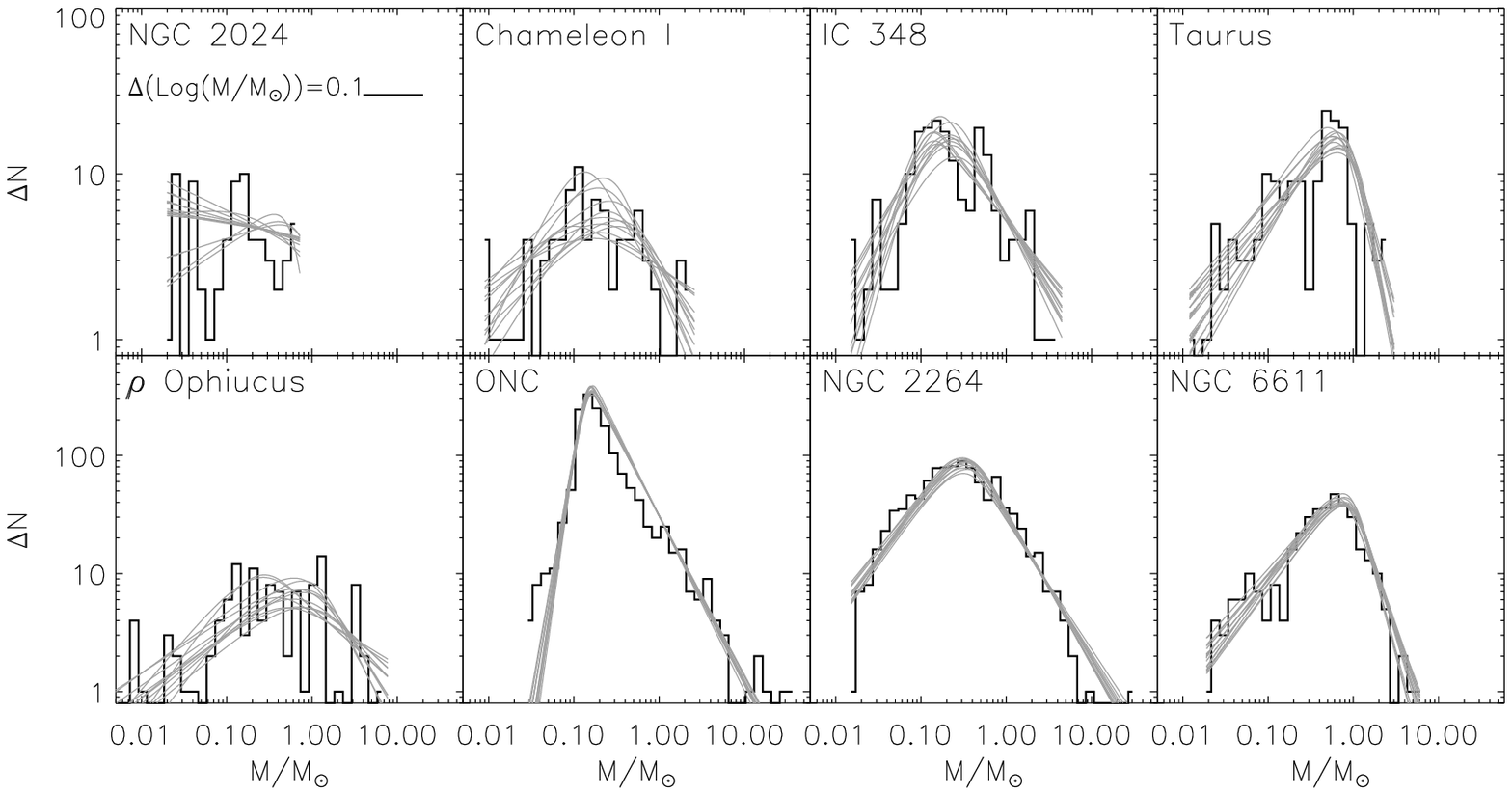}
\end{center}
\vspace{-3cm}
\caption{The grey lines show for each of the eight clusters considered in this work twelve random draws (selected randomly from the second half of the MCMC chain which contains in total $2\times 10^{5}$ iterations) of the IMF constructed using randomly selected triplets of the parameters ($\Gamma_{P}$, $\gamma_{P}$, $M_{P}$) which characterize the tapered power law mass function. These are compared (with the same normalisation) to the binned IMF of the clusters constructed by binning the individual stellar masses with a logarithmic bin size of ${\rm log}(M/{\rm M}_{\odot})=0.1$.}
\label{fig7}
\end{figure*}

\begin{figure*}
\centering
\begin{tabular}{cc}
\epsfig{file=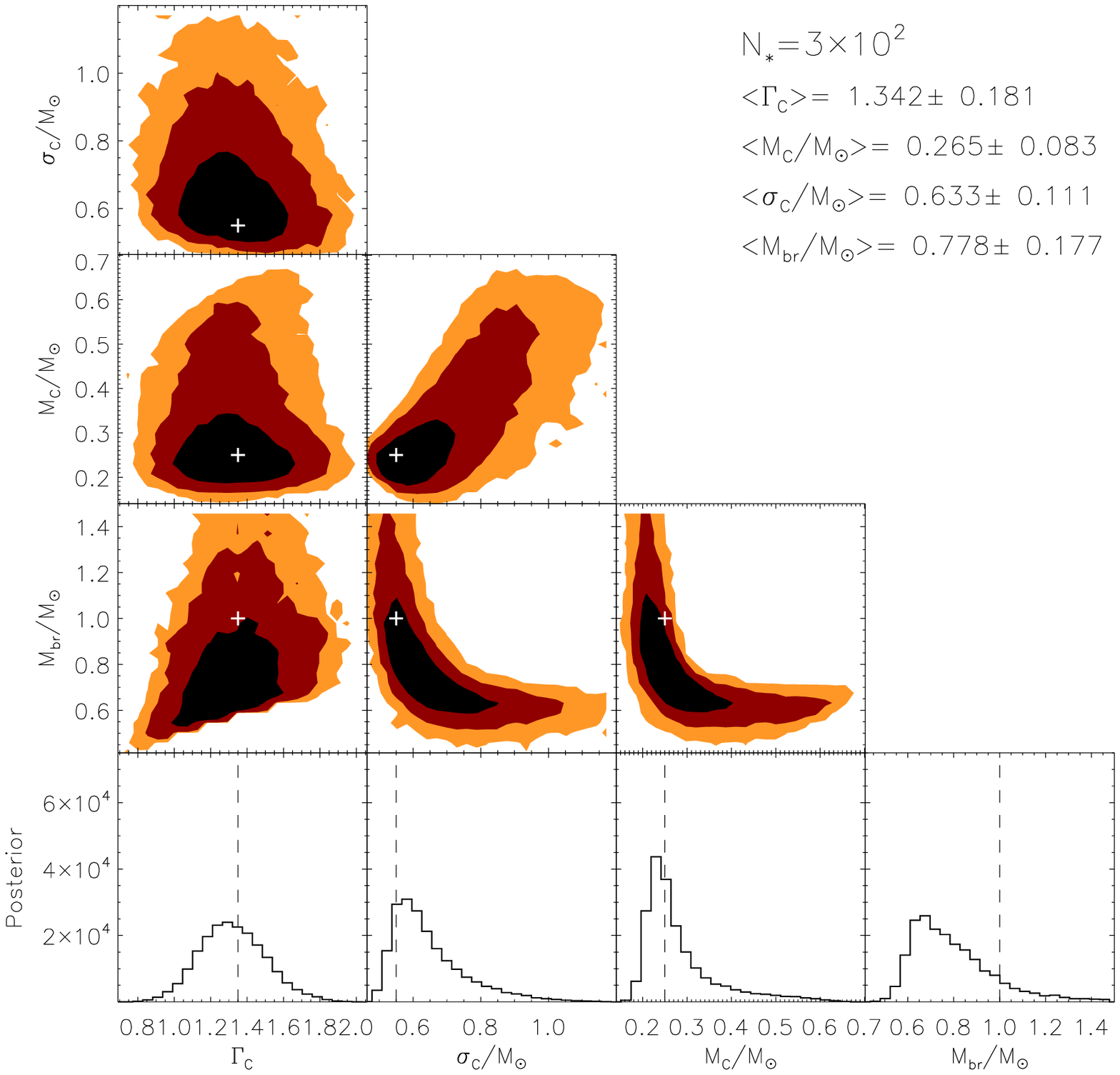,width=0.4\textwidth} &
\vspace{1cm}
\hspace{0.5cm}
\epsfig{file=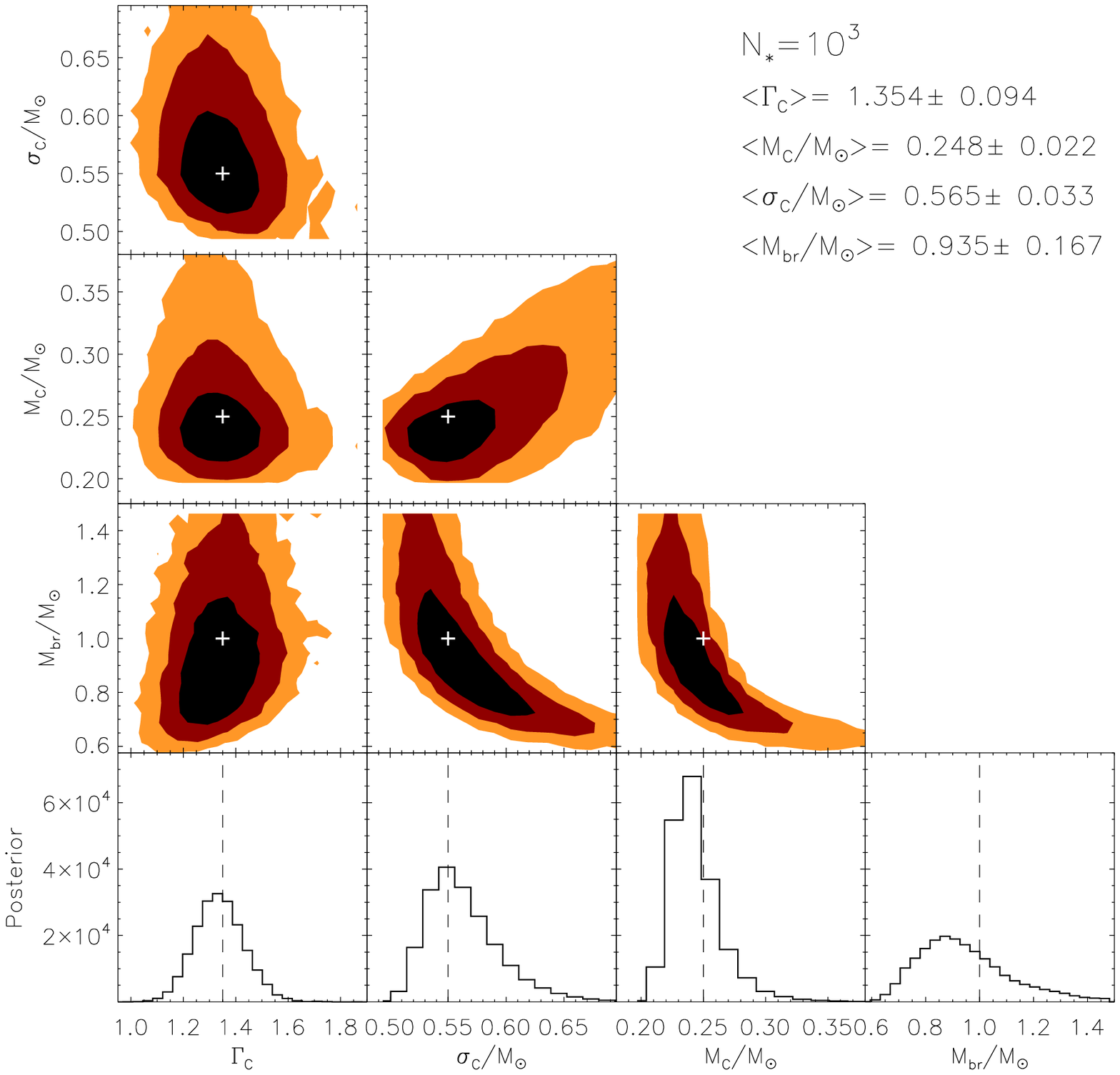,width=0.4\textwidth} \\
\epsfig{file=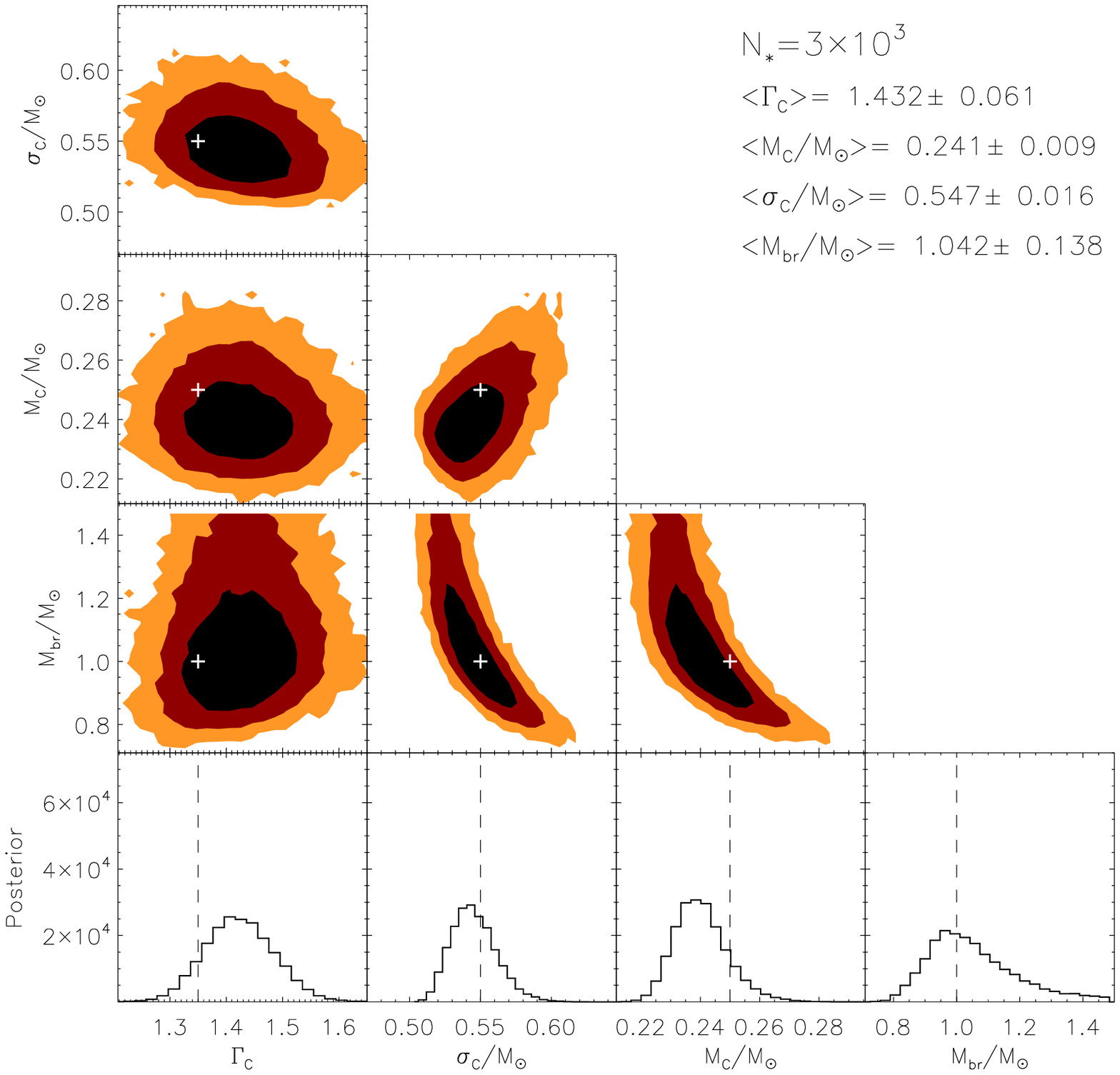,width=0.4\textwidth} &
\vspace{1cm}
\hspace{0.5cm}
\epsfig{file=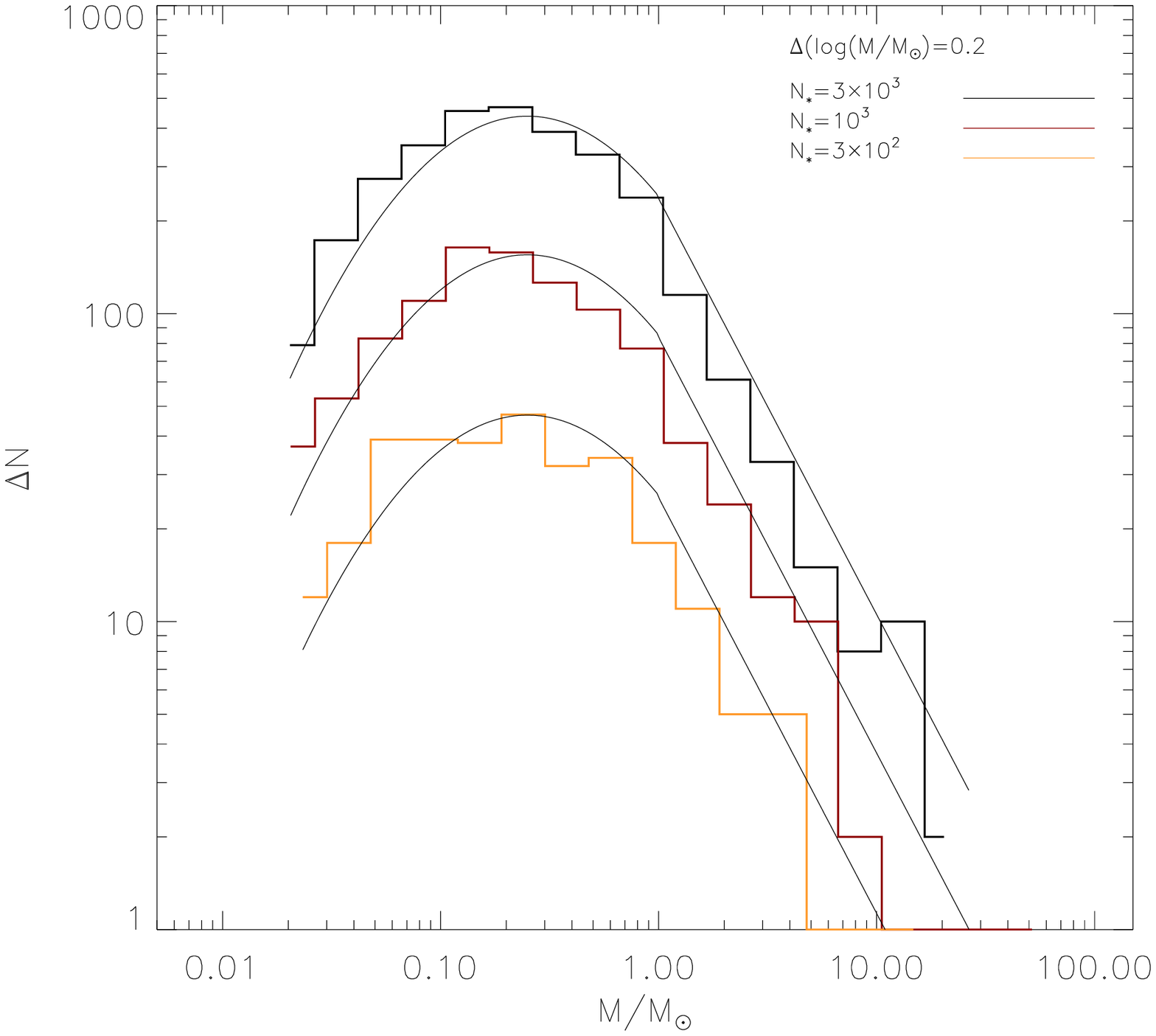,width=0.45\textwidth}
\end{tabular}
\caption{Posterior probability density functions (pPDFs, bottom row) and 2D density probability functions of the IMF parameters for three synthetic clusters with $N_{\star}=3\times10^{2}$ (top left), $N_{\star}=10^{3}$ (top right), and $N_{\star}=3\times10^{3}$ (bottom left) when the likelihood function is described by the lognormal+power law mass function (Eq.~\ref{eq2}). The ensemble of stars for which the IMF parameters have been inferred have been constructed using the lognormal+power law mass function with values of the parameters that are those derived for the Galactic field (Chabrier 2005) ($\Gamma_{C}=1.35$, $M_{C}=0.25$ M$_{\odot}$, $\sigma_{C}=0.55$ M$_{\odot}$, and $M_{br}=1$ M$_{\odot}$). The 1D pPDFs and the 2D density distributions of the parameters are binned in 25 bins. The latter are shown at the $1\sigma$ (black contours), $2\sigma$ (red contours), and $3\sigma$ (yellow contours) confidence levels. The IMF of these three synthetic clusters is shown in binned form (with ${\rm log}(M/{\rm M}_{\odot})=0.2$) in the lower right panel. Over-plotted to them is lognormal+power law function with the Galactic field values of the parameters. The Galactic field values of the parameters are shown with the white crosses in the 2D figures and with the black dashed lines in the 1D figures. The figure shows the ability of the Bayesian method with an MCMC approach in recovering the injected values of the parameters. The mean values of the parameters displayed in the upper right corner of each panel are shown for the sake of comparison only and are not necessarily representative of the model with the highest posterior probability.}
\label{fig8}
\end{figure*}

\begin{figure*}
\begin{center}
\includegraphics[width=0.9\textwidth,height=15cm]{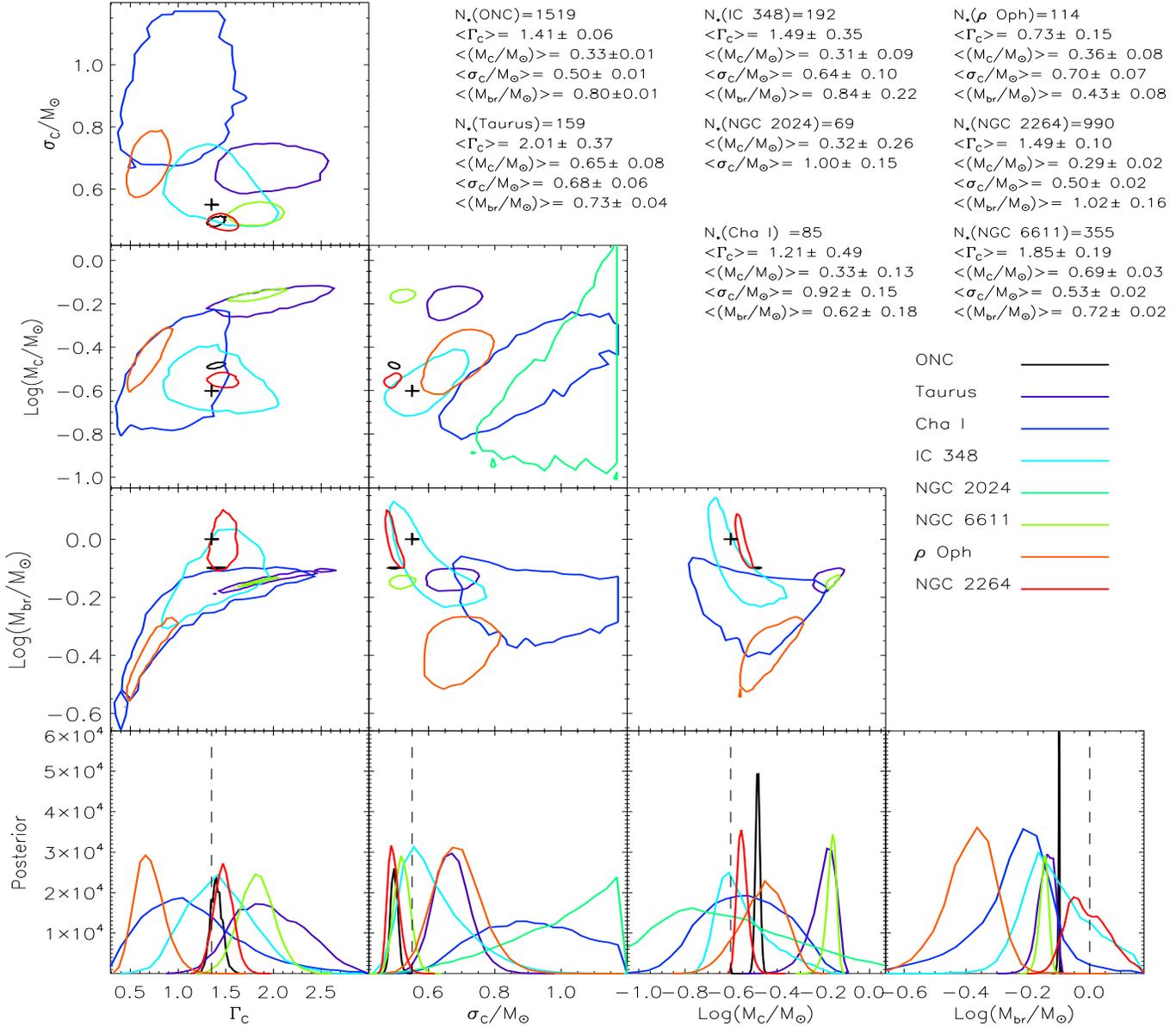}
\end{center}
\vspace{0.5cm}
\caption{Posterior probability density functions (pPDFs, bottom row) and 2D density probability functions of the parameters that describe the IMF for the eight young stellar clusters considered in this work when the likelihood function is given by lognormal+power law mass function (Eq.~\ref{eq2}). The four parameters of this mass function are: the width of the lognormal distribution $\sigma_{C}$, the characteristic mass $M_{C}$, the slope of the power law at the intermediate- to high mass end $\Gamma_{C}$, and the break point that marks the transition between the lognormal and power law components, $M_{br}$. The 2D posterior probability densities are are shown at the $1\sigma$ level. The Galactic field values of the parameters derived by Chabrier (2005) for the Galactic field IMF are shown with the black crosses in the 2D figures and with the black dashed lines in the 1D figures. The mean values of the parameters displayed in the upper right corner of the figure are shown for the sake of comparison only and are not necessarily representative of the model with the highest posterior probability.}
\label{fig9}
\end{figure*}

\begin{figure*}
\begin{center}
\includegraphics[width=0.9\textwidth]{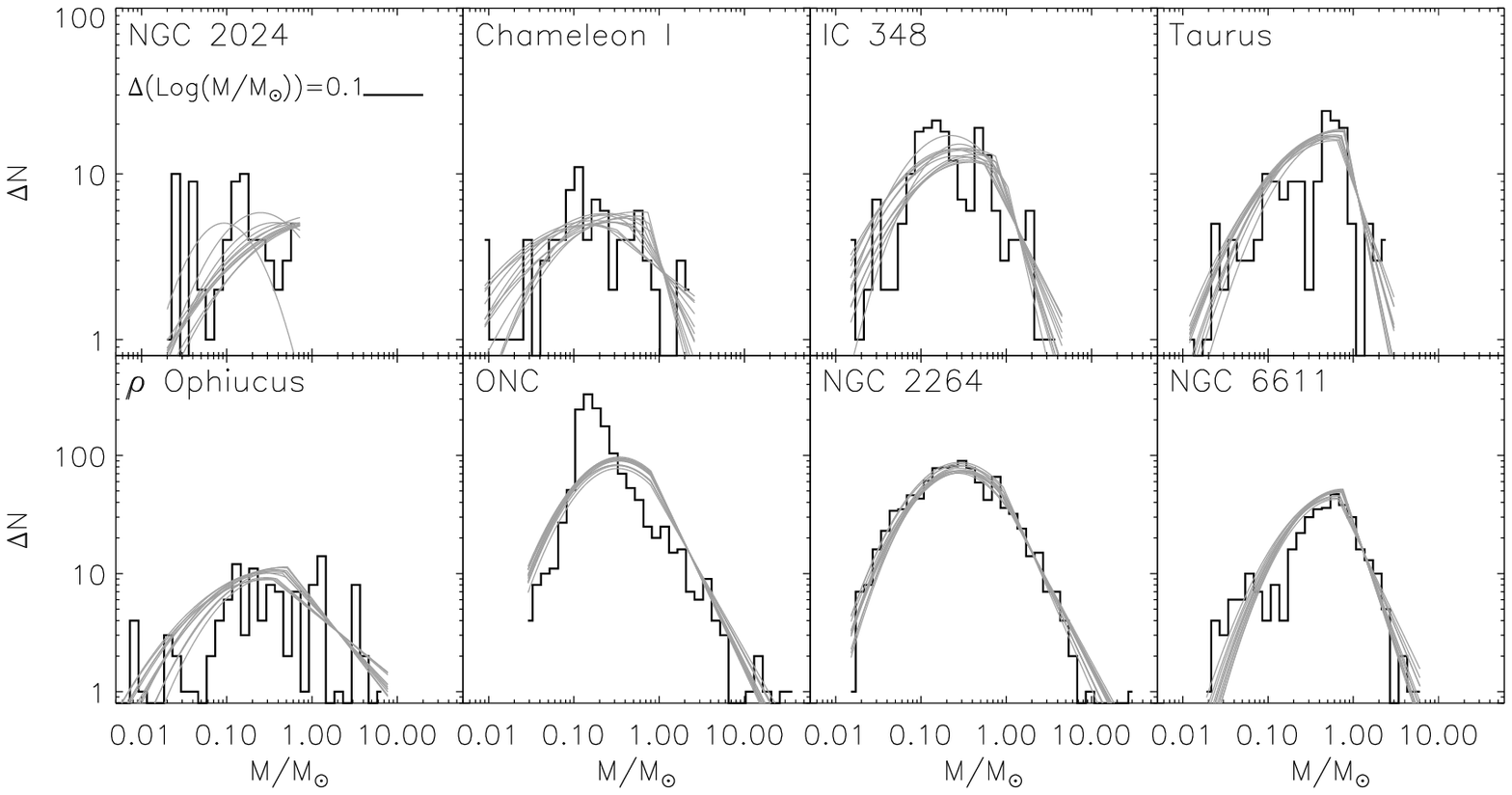}
\end{center}
\vspace{-3cm}
\caption{The grey lines show for each of the eight clusters considered in this work twelve random draws (selected randomly from the second half of the MCMC chain which contains in total $2\times 10^{5}$ iterations) of the IMF constructed using randomly selected quadruplets of the parameters ($\Gamma_{C}$, $\sigma_{C}$, $M_{C}$, $M_{br}$) which characterize the lognormal+power law function. These are compared (with the same normalization) to the binned IMF of the clusters constructed by binning the individual stellar masses with a logarithmic bin size of ${\rm log}(M/{\rm M}_{\odot})=0.1$.}
\label{fig10}
\end{figure*}

\begin{figure*}
\centering
\begin{tabular}{cc}
\epsfig{file=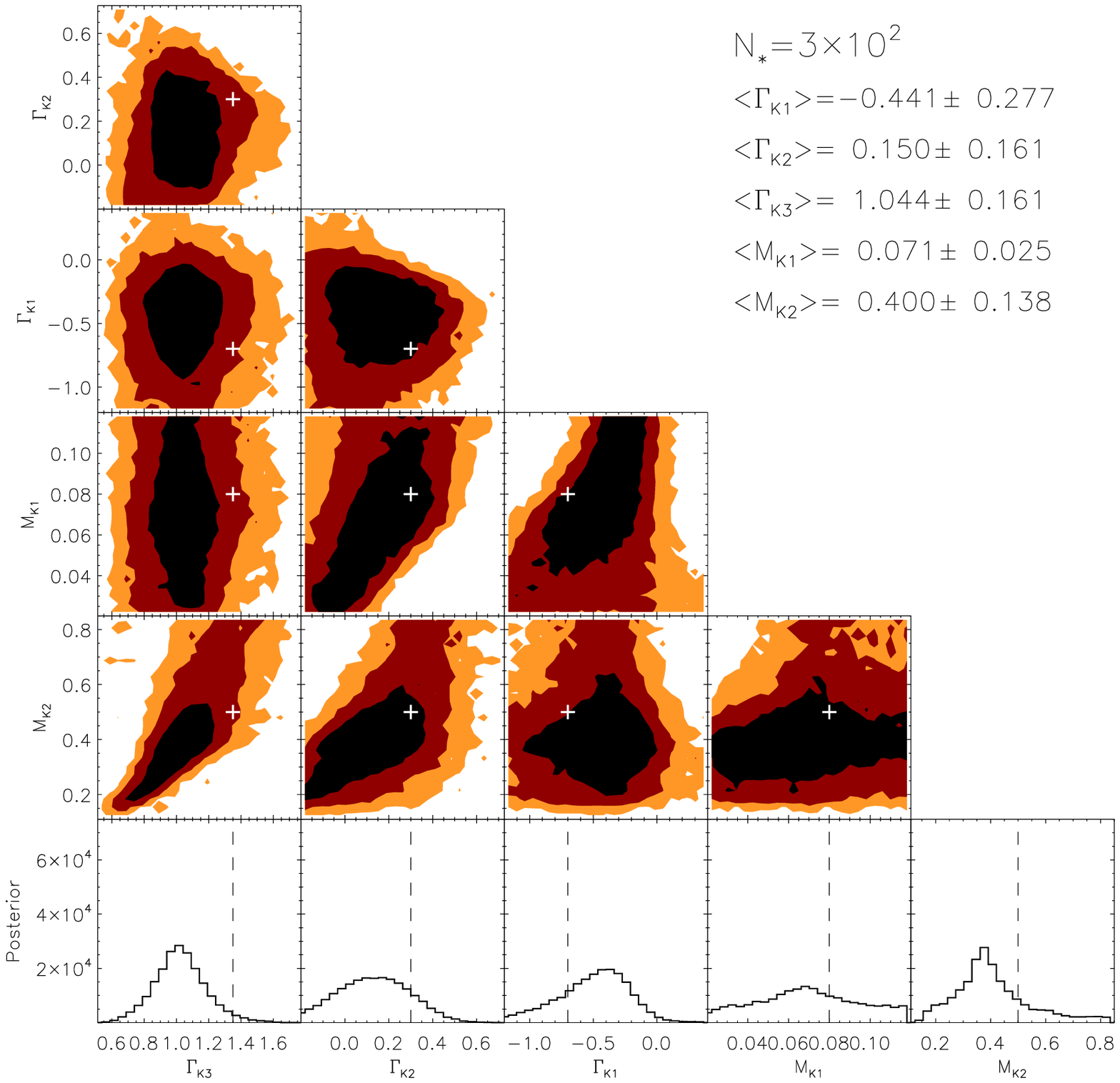,width=0.4\textwidth} &
\vspace{1cm}
\hspace{0.5cm}
\epsfig{file=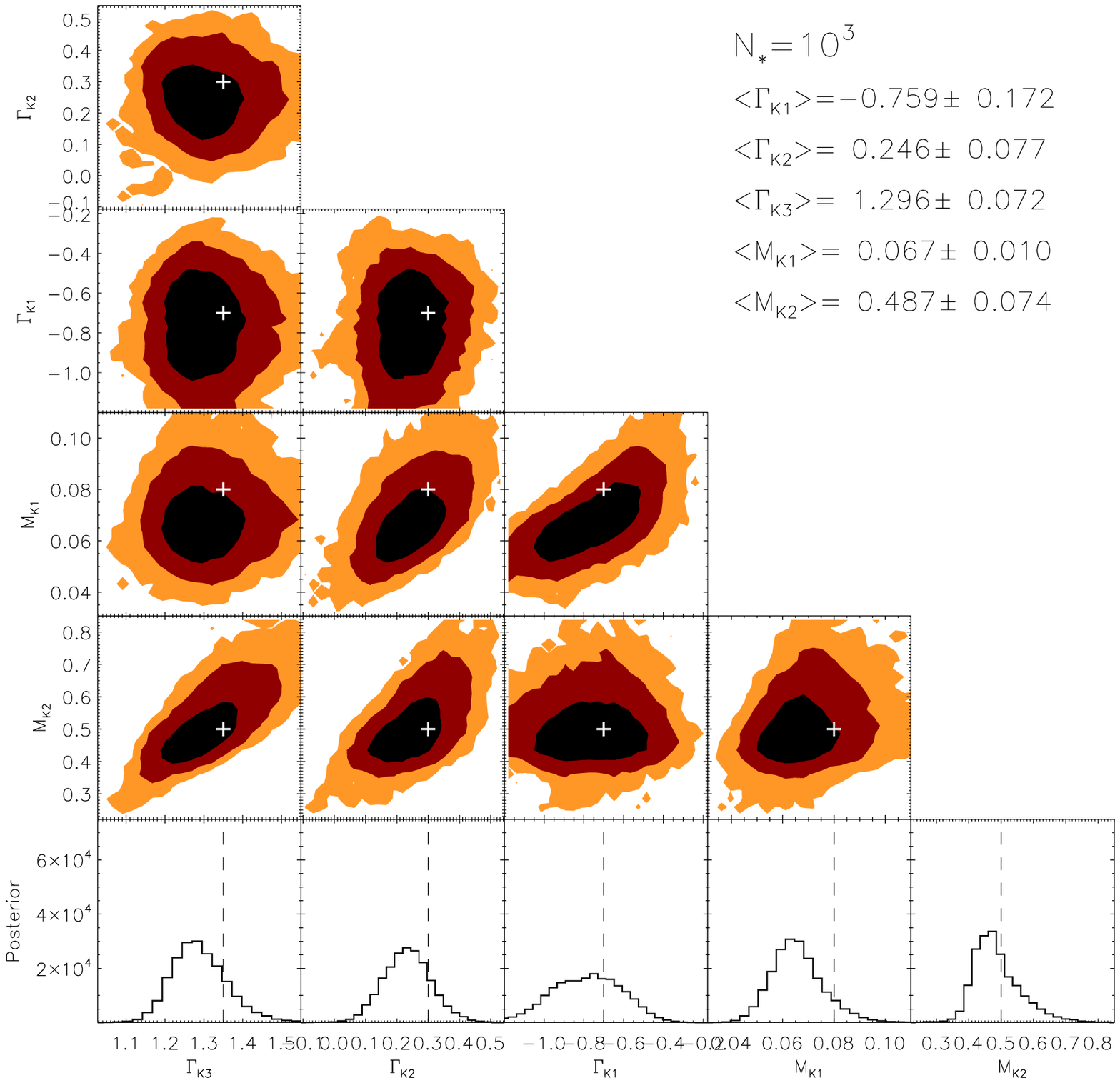,width=0.4\textwidth} \\
\epsfig{file=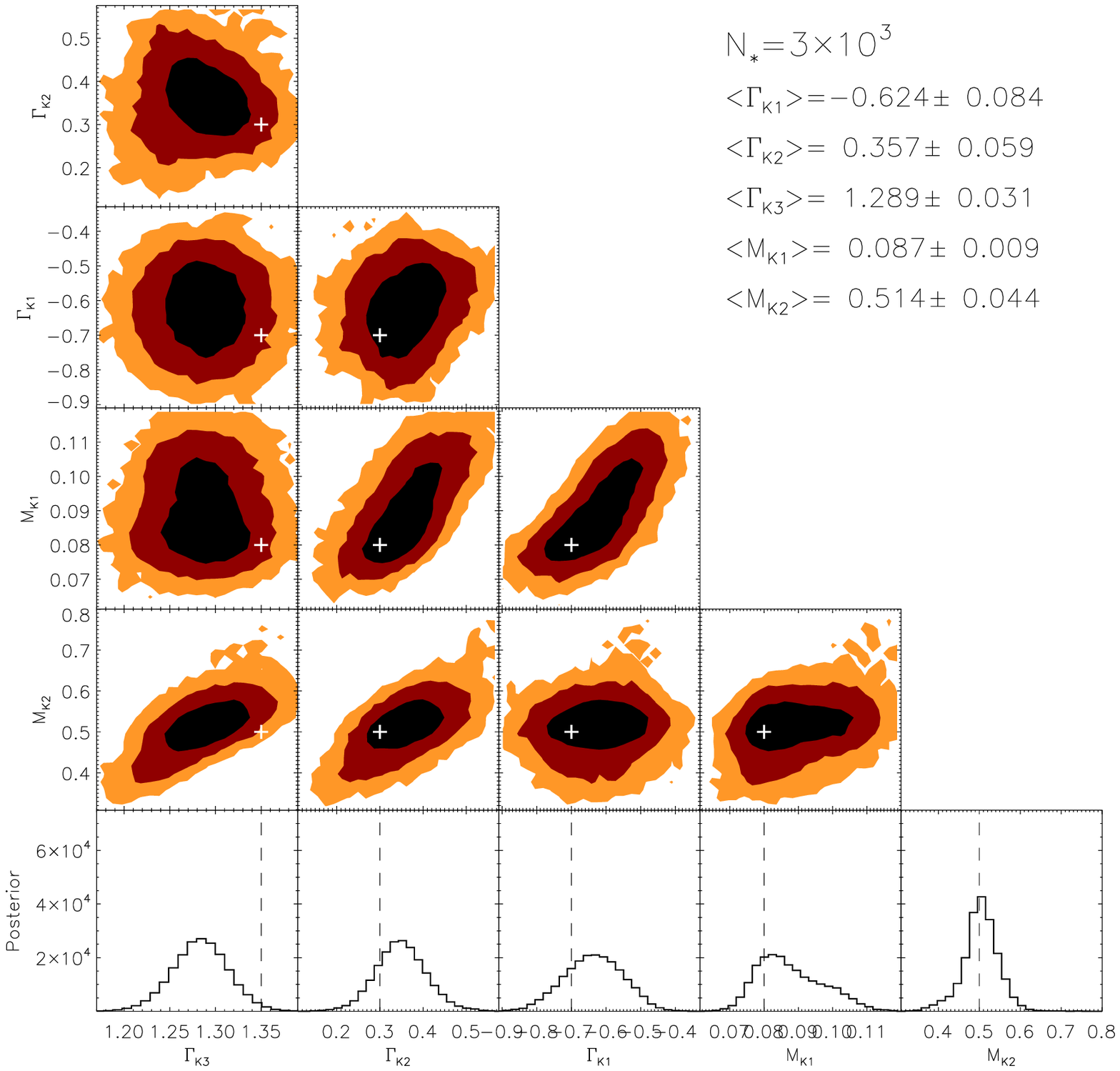,width=0.4\textwidth} &
\vspace{1cm}
\hspace{0.5cm}
\epsfig{file=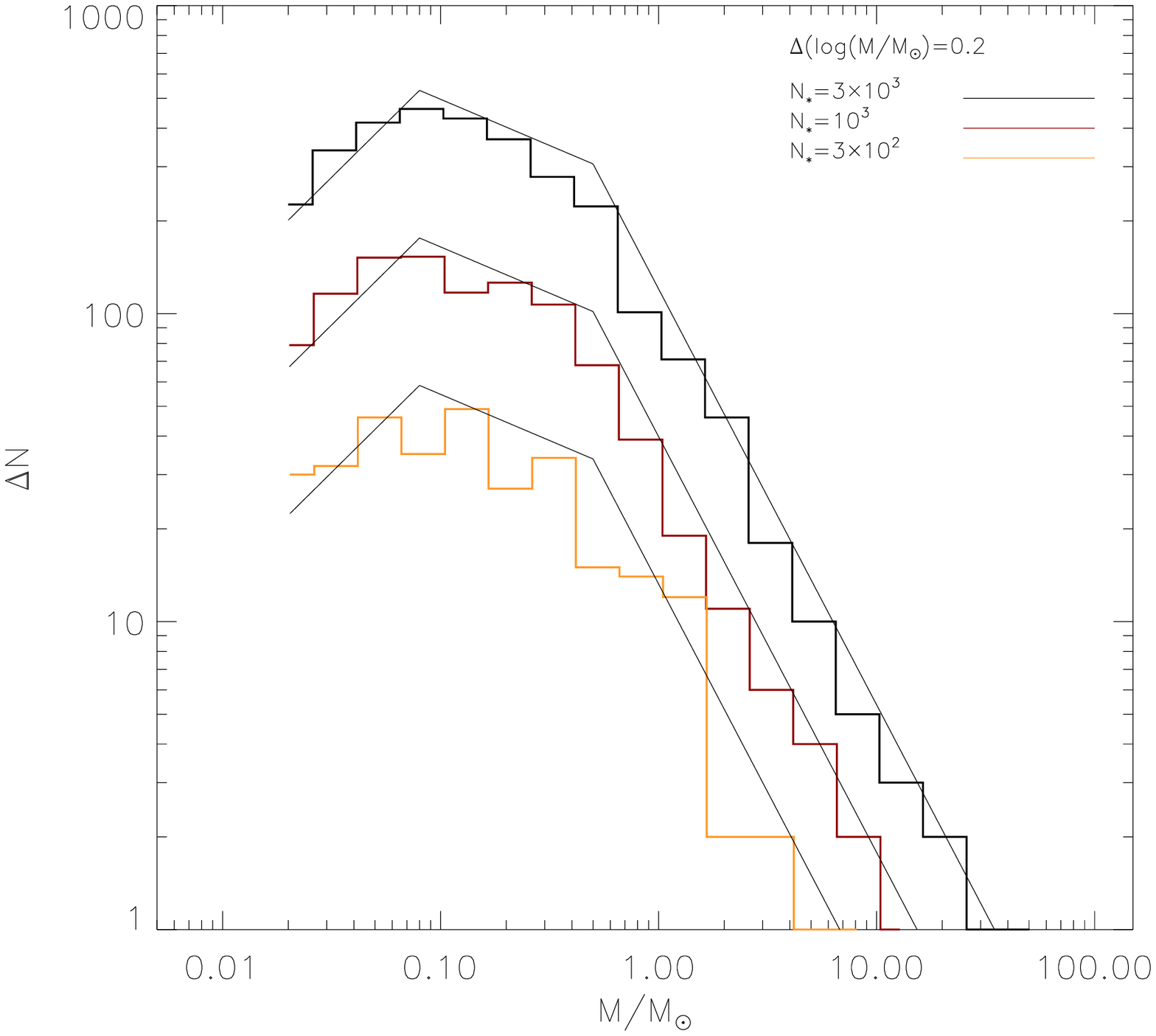,width=0.45\textwidth}
\end{tabular}
\caption{Posterior probability density functions (pPDFs, bottom row) and 2D density probability functions of the IMF parameters for three synthetic clusters with $N_{\star}=3\times10^{2}$ (top left), $N_{\star}=10^{3}$ (top right), and $N_{\star}=3\times10^{3}$ (bottom left) when the likelihood function is described by a three-component power law function (Eq.~\ref{eq3}). The ensemble of stars for which the IMF parameters have been inferred have been constructed using the three-component power law mass function with values of the parameters that are those derived for the Galactic field (Kroupa 2002) ($\Gamma_{K1}=-0.7$, $\Gamma_{K2}=0.3$, $\Gamma_{K3}=1.35$, $M_{K1}=0.08$ M$_{\odot}$, and $M_{K2}=0.5$ M$_{\odot}$). The 1D pPDFs and the 2D density distributions of the parameters are binned in 25 bins. The latter are shown at the $1\sigma$ (black contours), $2\sigma$ (red contours), and $3\sigma$ (yellow contours) confidence levels. The IMF of these three synthetic clusters is shown in binned form (with ${\rm log}(M/{\rm M}_{\odot})=0.2$) in the lower right panel. Over-plotted to them is lognormal+power law function with the Galactic field values of the parameters. The Galactic field values of the parameters are shown with the white crosses in the 2D figures and with the black dashed lines in the 1D figures. The figure shows the ability of the Bayesian method with an MCMC approach in recovering the injected values of the parameters. The mean values of the parameters displayed in the upper right corner of each panel are shown for the sake of comparison only and are not necessarily representative of the model with the highest posterior probability.}
\label{fig11}
\end{figure*}

\begin{figure*}
\begin{center}
\includegraphics[width=0.9\textwidth,height=15cm]{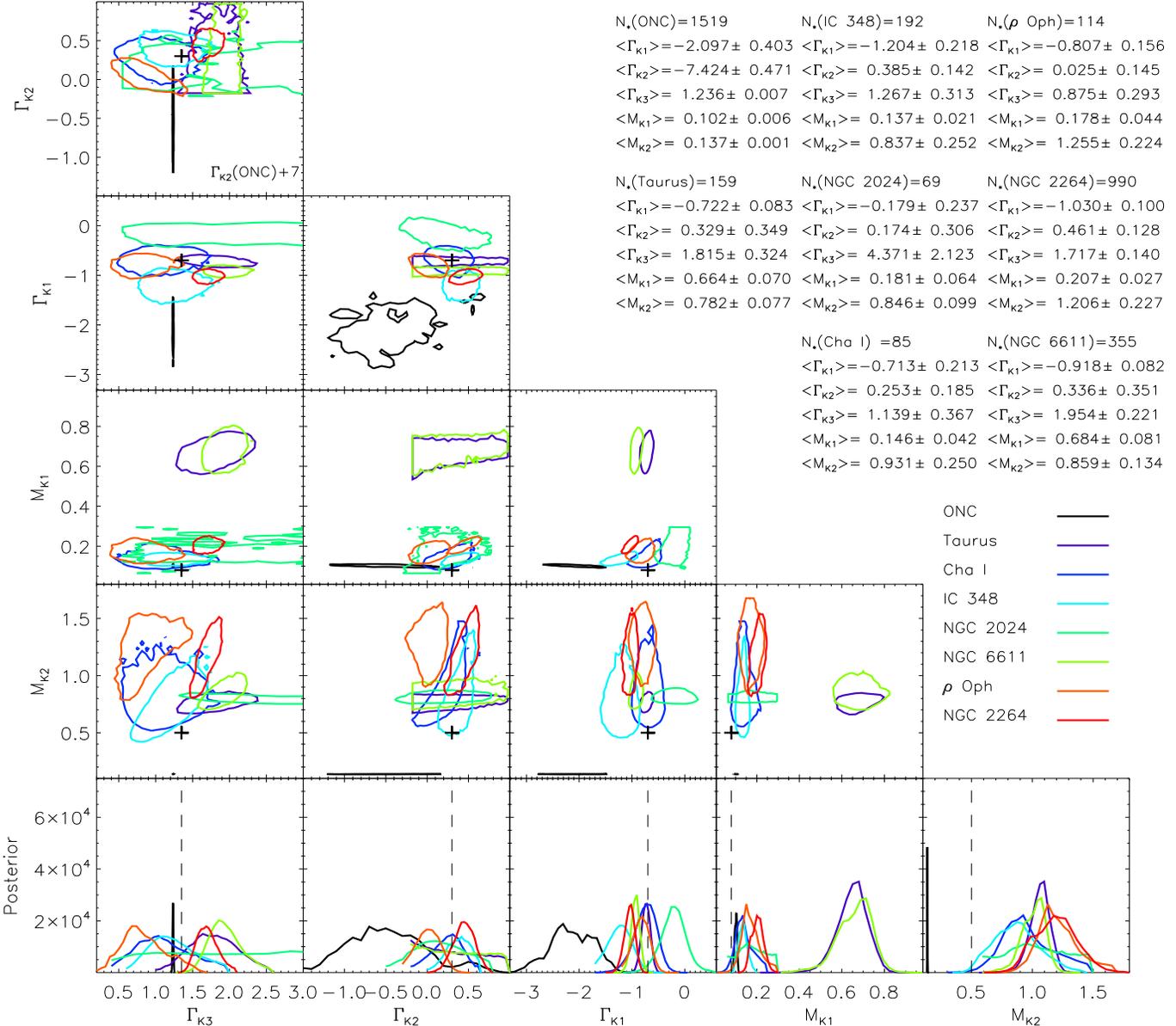}
\end{center}
\vspace{0.5cm}
\caption{Posterior probability density functions (pPDFs, bottom row) and 2D density probability functions of the parameters that describe the IMF for the eight young stellar clusters considered in this work when the likelihood function is given by three-component power law mass function (Eq.~\ref{eq3}). The five parameters of this mass function are the three slopes in the three mass ranges ($\Gamma_{K1}$, $\Gamma_{K2}$, and $\Gamma_{K3}$ when going from the low to the high mass regime, respectively) and the two break points which mark the transition between the three components of the mass functions ($M_{K1}$ and $M_{K2}$). The 2D posterior probability densities are are shown at the $1\sigma$ level. The Galactic field values of the parameters derived by Kroupa (2002) for the Galactic field IMF are shown with the black crosses in the 2D figures and with the black dashed lines in the 1D figures. The mean values of the parameters displayed in the upper right corner of the figure are shown for the sake of comparison only and are not necessarily representative of the model with the highest posterior probability.}
\label{fig12}
\end{figure*}

\begin{figure*}
\begin{center}
\includegraphics[width=0.9\textwidth]{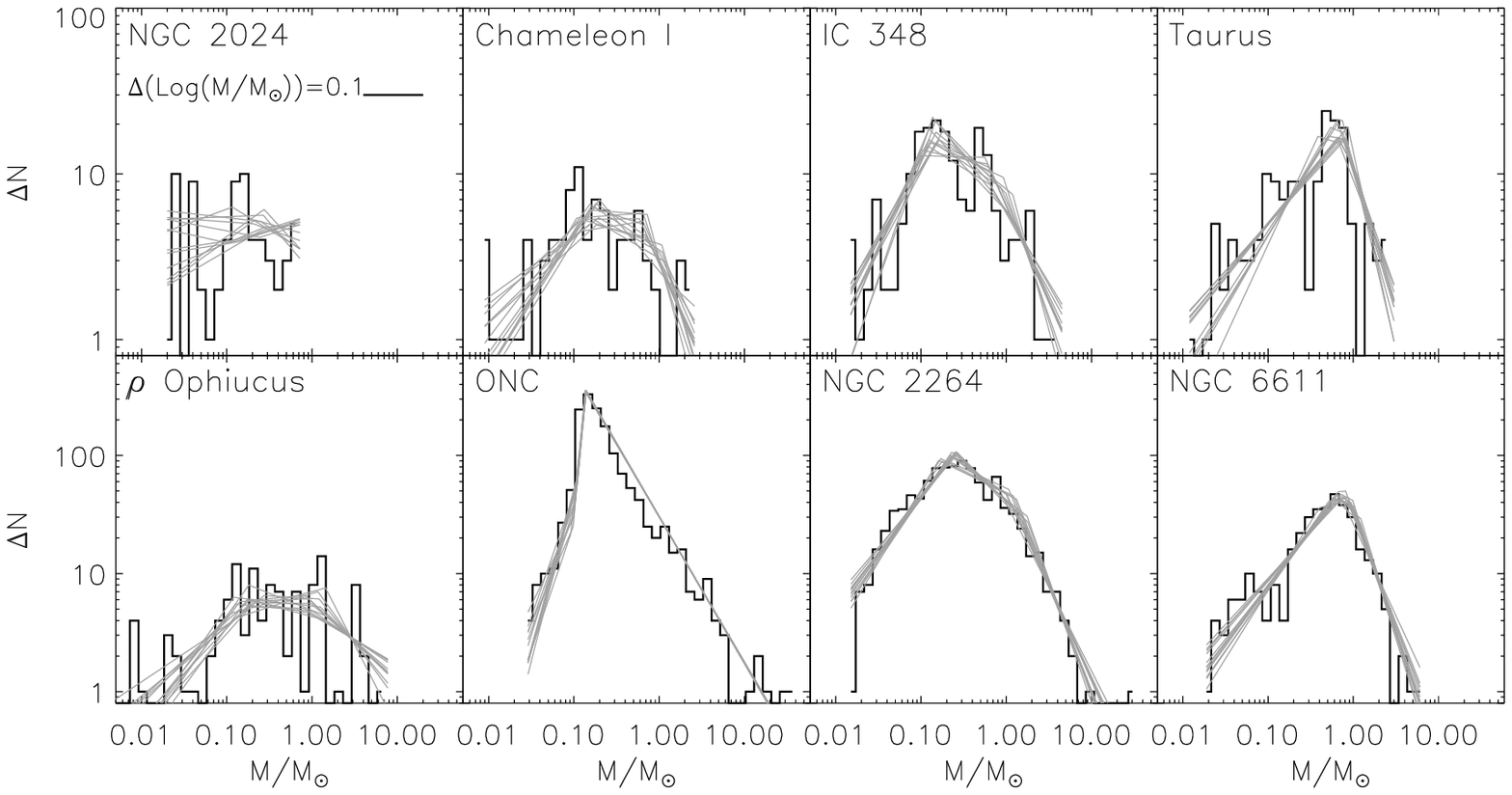}
\end{center}
\vspace{-3cm}
\caption{The grey lines show for each of the eight clusters considered in this work twelve random draws (selected randomly from the second half of the MCMC chain which contains in total $2\times 10^{5}$ iterations) of the IMF constructed using randomly selected quadruplets of the parameters ($\Gamma_{C}$, $\sigma_{C}$, $M_{C}$, $M_{br}$) which characterize the three-component power law function. These are compared (with the same normalization) to the binned IMF of the clusters constructed by binning the individual stellar masses with a logarithmic bin size of ${\rm log}(M/{\rm M}_{\odot})=0.1$.}
\label{fig13}
\end{figure*}

\noindent {\it Orion Nebula Cluster}: The Orion Nebula Cluster (ONC) located at a distance of $\approx 400$ pc from the sun, is the nearest cluster which hosts a number of massive B and O stars. A census of the stellar content of the ONC has been performed by several groups (Hillenbrand 1997; Palla \& Stahler 1999; Hillenbrand \& Carpenter 2000; Slesnick et al. 2004; Andersen et al. 2011). In this work, we use the most recent determination of stellar masses in the ONC obtained by Da Rio et al. (2012b). These authors used the Wide Field Imager (WFI) located at the 2.2 m MPG/ESO telescope at La Silla. The field of view of the WFI is large enough ($\gtrsim 30' \times 30'$) to encompass the entire ONC in one pointing. Their photometry included measurement in the I-band and two medium band filters at $\lambda=753$ and $770$ nm.  From the colours obtained in these three bands, they were able to derive the effective temperatures of 1750 sources and place them on the Hertzsprung-Russel (HR) diagram. The masses (and ages) were derived using the evolutionary tracks of Baraffe et al. (1998) and D'Antona \& Mazzitelli (1998). Since Da Rio et al. (2012b) could not assign masses and ages for about 25 \% of the sources when using the Baraffe et al. tracks as these are located above the 1 Myrs isochrone, we use in the work the masses they have derived using the D'Antona \& Mazzitelli (1998) evolutionary tracks. We complemented the Da Rio et al. (2012b) data by the census of massive stars (in total 27 stars missing from the Da Rio et al. 2012b data) in the ONC obtained by Hillenbrand (1997). Da Rio et al. (2012b) reported the presence of a contamination by a background population in the ONC and therefore, we discard stars that have a membership probability of belonging to the ONC that is $< 99\%$. The total remaining sample is composed of $N_{\star}$(ONC)$=1519$. The mass range of stars in this sample is [0.029-45.7] M$_{\odot}$. Hillenbrand et al. (2013) used spectroscopic measurements and derived the spectral type and stellar masses for 619 ONC stars. Out of these, the spectral types for 437 stars were compared to the photometric spectral types of Da Rio et al. (2012b).  As discussed in Hillenbrand et al. (2013, section 5.2 and Fig. 10 in their paper), the scatter between the spectroscopically determined spectral type and the ones determined by Da Rio et al. (2012b) is of the order of 1.75 spectral sub-class. These variations are small enough such as not to induce any significant variation in the derived stellar masses.  It is important to note that the census of stars in the ONC is also very likely contaminated by a significant foreground population as recently shown and discussed by Alves \& Bouy (2012) and Bouy et al. (2014). 
\newline
\noindent {\it $\rho$ Ophiuchi}: Several groups have investigated over the past two decades the census of stars in this young star forming region (e.g., Rieke et al. 1989; Luhman \& Rieke 1999; Williams et al. 1995; Erickson et al. 2011). Using WIRCam, Alves de Oliveira et al. (2010) performed a comprehensive census of the stellar content in $\rho$ Ophiuchi. They confirmed many of their sources using a follow-up spectroscopic study and complemented it with spectroscopically confirmed sources from studies by other groups (Wilking et al. 2008; McClure et al. 2010; Geers et al. 2010; Erickson et al. 2011). In total, 250 spectroscopically confirmed members of the cluster were placed on the H-R diagram (Alves de Oliveira et al. 2012). In this work, we make use of the sample of sources that is complete (with $A_{V} \le 8$) and with this choice, the number of sources is reduced to $N_{\star}$($\rho$ Oph)$=114$. Stellar masses were derived using the Baraffe et al. (1998) and Siess et al. (2000) evolutionary tracks and assuming a distance to the cluster of 130 pc. The masses of the stars considered in this sample extend over the mass range [0.033-7.22] M$_{\odot}$.      	
\newline
\noindent {\it Taurus}: A census of the protostellar and stellar population in the Taurus star forming region has been performed by Luhman (2000), Brice$\tilde {\rm n}$o et al. (2002), Luhman et al. (2003b), Luhman (2004a), and Scelsi et al (2007,2008). The most recent available information concerning this cluster is provided by Luhman et al. (2009). These authors performed spectroscopy on infrared sources that have been identified in the clouds with the XMM-{\it Newton} Extended Survey of the Taurus Molecular Cloud (XEST, G\"{u}del et al. 2007) and the Taurus Spitzer Survey (Rebull et al. 2010). The masses and ages of members of Taurus were estimated using the theoretical evolutionary tracks of Chabrier et al. (2000) for M$_{\star}$/M$_{\odot} \leq 0.1$, the Baraffe et al. (1998) tracks for $0.1 <$ M$_{\star}$/M$_{\odot} \leq 1$, and Palla \& Stahler (1999) for M$_{\star}$/M$_{\odot} > 1$. Available proper motion observations were also used by Luhman et al. (2009) in order to put additional constraints on the membership of the sources to the Taurus cluster. In total, the newest census of Luhman et al. (2009) contain $N_{\star} {\rm (Taurus)}=159$ spanning the mass range [0.012-3] M$_{\odot}$. 
\newline
\noindent {\it Cha I}: The stellar content of the Chamaleon I star forming region (Cha I) has been investigated by Mamajek et al. (2000), Song et al (2004), Luhman et al. (2004b), Lyo et al. (2006), and Mu$\check{\rm z}$i\'{c} et al. (2011). Using optical and near-infrared photometry with follow up spectroscopy, Luhman (2007) presented a census of the stellar population of Cha I (226 sources). Out of these 226 known members, 11 were excluded due to very uncertain spectral types. In order to ensure for a better completeness, Luhman (2007) imposed a limit of $A_{J} \leq 1.2$ and $A_{J} \leq 1.4$ for the two distinct stellar concentrations/clusters in the Cha I star forming regions (Cha I South and Cha I North, respectively). This reduced the number of known members to 85 and 34 in Cha I South and Cha I North, respectively. The masses of the stars were derived, as for Taurus, using the stellar evolutionary tracks of Chabrier et al. (2000) for M$_{\star}$/M$_{\odot} \leq 0.1$, Baraffe et al. (1998) for  $1 <$ M$_{\star}$/M$_{\odot} \leq 1$ and Palla \& Stahler (1999) for masses M$_{\star}$/M$_{\odot} > 1$. In this work, we consider the population of Cha I South. The masses of the stars found in Cha I South fall in the range [0.009-2.6] M$_{\odot}$.       
\newline
\noindent {\it IC 348}: The IC 348 cluster is located at a distance of 310 pc and its estimated age is $2-3$ Myr (e.g., Stelzer et al. 2012). Its stellar content has been investigated by Lada \& Lada (1995), Najita et al. (2000), Muench et al. (2003), Preibisch et al. (2003), Burgess et al. (2009), and Alves de Oliveira et al. (2013). In this work, we use the stellar census obtained by Luhman et al. (2003a). Cluster membership was determined by placing the stars (268) on an extinction corrected diagram of $I-K_{s}$ versus $H$ and $I-Z$ versus $H$. Spectra were obtained for these 268 stars. As for Cha I and Taurus, Luhman et al. (2003a) used the stellar evolutionary tracks of Chabrier et al. (2000), Baraffe et al. (1998), and Palla \& Stahler (1999) to derive the masses and ages of the stars for masses in the ranges ${\rm M}_{\star}/{\rm M}_{\odot} \leq 0.1$, $0.1 < {\rm M}_{\star}/{\rm M}_{\odot} \leq 1$ and ${\rm M}_{\star}/{\rm M}_{\odot} >1$, respectively. Luhman et al. (2003a) then defined an IMF-sample of stars by selecting only members that have $A_{V} \leq 4$. This reduced the number of members to $N_{\star}$ (IC 348)$=192$, which is the sample of stars we adopt in this work. The masses of the stars in this sample extend over the range [$0.015-4.5$] M$_{\odot}$. 
\newline
\noindent {\it NGC 2264}:  Investigations of the stellar content of the young (2-4 Myrs) cluster NGC 2264 has been performed by Flaccomio et al. (1999), Park et al. (2000), and Sung et al. (2004). Sung et al. (2008,2010) measured the masses of stars in NGC 2264 from photometric observations. They constructed a color-magnitude diagram ($I_{c}$ versus $V-I_{c}$) of cluster members and derived their masses by comparing them to the stellar evolutionary models of Baraffe et al. (1998) and Siess et al. (2000). They measured the masses of stars in the locus of the pre-main sequence (PMS) of NGC 2264, using a mean reddening function. Heavily embedded stars and stars below the pre-main sequence were excluded. This reduced the sample of cluster members with which Sung et al. (2012) constructed the (binned) IMF of NGC 2264 to $N_{\star}$(NGC 2264)$=990$. This is the sample of stars in NGC 2264 that is considered in this work. The masses of the stars considered in this sample span over the range [$0.015-33.8$] M$_{\odot}$. 
 \newline
\noindent {\it NGC 6611}: The stellar content and the IMF of NGC 6611 have been investigated by several groups (e.g., Walker 1961, Sagar 1979, Bonatto 2006). In this work, we use the sample of cluster members whose masses have been derived by Oliveira et al. (2009) for the low mass populations in the cluster using observations from the {\it Hubble Space Telescope}. This is complemented by the census of more massive stars derived from ground-based observations by Oliveira et al. (2005). Oliveira et al. (2009) constructed the CMD of the cluster and derived two distinct samples of masses for the cluster members by assuming ages of 2 and 3 Myrs. These were derived using the 2 and 3 Myrs isochrones of Baraffe et al. (1998) (for lower masses) and Siess et al. (2000) (for higher stellar masses). In this work, we have considered a single sample of $N_{\star}$(NGC 6611)$=355$ stars and have calculated the mass of each star as being the mean of the 2 and 3 Myrs mass estimates.  
\newline
\noindent {\it NGC 2024}: NGC 2024 is a young, very embedded cluster (e.g., Comeron et al. 1996; Haisch et al. 2000). The only publicly available and documented catalogue of members of NGC 2024 is presented in Levine et al. (2006). The sample of stars in this catalogue is however restricted to the mass range [0.02-0.72] M$_{\odot}$. Levine et al. (2006) combined near infrared spectroscopic observations with $JHK$ photometry. They derived the stellar masses and ages of the stars from the H-R diagram using the Baraffe et al. (1998) evolutionary tracks. Despite the fact that the data of NGC 2024 is limited to the substellar regime, we still include it in our sample of clusters and take it as an example of the effects of "missing data" on the inference of the parameters that describe the underlying IMF.     

Fig.~\ref{fig1} and Fig.~\ref{fig2} display a binned form of the IMF for each of these cluster. The bins are of equal logarithmic sizes of ${\rm log} (M/{\rm M_{\odot}})=0.1$ (red lines) and ${\rm log} (M/{\rm M_{\odot}})=0.2$ (yellow lines). Overlaid to the IMF of each cluster and for each choice of the logarithmic mass bin size are the Galactic field star IMF of Parravano et al. (2011) (bottom panels), Chabrier (2005) (middle panels), and Kroupa (2002) (top panels) with their fiducial parameters given in \S.~\ref{models}. 
  
\section{BAYESIAN ANALYSIS AND ITS IMPORTANCE}\label{bayesian}

By performing only a visual inspection of the mass functions of the eight clusters displayed in Fig.~\ref{fig1} and Fig.~\ref{fig2}, it is difficult to assess any variations in the parameters that describe their underlying IMF. It is also equally difficult to assess which IMF model is a better description for the underlying IMF of each individual cluster. The classical method that is commonly employed to infer the parameters that describe the shape of the IMF is a minimization of the $\chi^{2}$ between the data and and IMF model. As pointed by several authors (Ma\'{i}z Apell\'{a}niz \& \'{U}beda 2005; Weisz et al. 2013), both the values of the parameters describing the underlying IMF of a stellar cluster as well as the uncertainties on these parameters are dependent on the choice of bin size. An additional complication is created by the somewhat arbitrary and subjective choice of the mass range over which a functional form of the IMF is fitted, particularly in the presence of break points such as in the case of the Chabrier and Kroupa IMFs. The dependence of the derived parameters of the IMF on the bin size is further illustrated in Fig.~\ref{fig3} and Fig.~\ref{fig4}. Fig.~\ref{fig3} displays binned IMFs of 3 clusters with $N_{\star}=10^{2}$, $10^{3}$, and $10^{4}$ (yellow, red, and black respectively) and where the masses, which are randomly sampled from the Parravano et al. IMF (Eq.~\ref{eq1}) in the mass range [0.02-150] M$_{\odot}$, are binned using equals size logarithmic mass bins of $\Delta {\rm log} (M/M_{\odot})=0.1$ (left panel), $0.2$ (middle panel), and $0.3$ (right panel). The binned mass functions are then fitted with the tapered power law IMF (De Marchi et al. 2010; Parravano et al. 2011). The parameters $\Gamma_{P}$, $M_{P}$, and $\gamma_{P}$ are derived by minimizing the $\chi^{2}$ between the fit function and the binned IMFs using a Levenberg-Marquart algorithm. Fig.~\ref{fig4} displays the $1\sigma$ uncertainty, in the form of a percentage, on $\Gamma_{P}$ (top panel), $M_{P}$ (middle panel), and $\gamma_{P}$ (lower panel) as a function of the logarithmic bin size (in the range  $0.05 \leq \Delta {\rm log} (M/{\rm M}_{\odot)} \leq 0.4$), and for the three clusters with $N_{\star}=10^{2}$ (yellow), $10^{3}$ (red), and $10^{4}$ (black). The figure clearly shows that the uncertainty on the derived parameters is bin-dependent and that it increases with the increasing  size of the bin. The use of Bayesian statistics for the inference of the parameters that describe the underlying shape of the IMF of a given stellar cluster with resolved populations makes it possible to avoid the pitfalls that can be caused by these subjective choices.      

In Bayesian statistics, assessing the probability that a specific model ${\cal{M}}_{i}$  out of a set of possible models $\cal{M}$ (i.e., a specific combination of the model's parameters) can explain a set of data $\cal{D}$ is given by the fundamental equation (i.e., Bayes' theorem; Bayes 1763):

\begin{equation}
P\left({\cal M}_{i}|{\cal D}\right)=\frac{P\left({\cal M}_{i}\right)~P\left({\cal D}|{\cal M}_{i}\right)} {P(\cal{D})},
\label{eq4}
\end{equation}

\noindent where $P\left({\cal M}_{i}|{\cal D}\right)$ is the posterior probability (hereafter the {\it posterior}) that the model ${\cal M}_{i}$ is correct given the data ${\cal D}$, $P\left({\cal M}_{i}\right)$ is the prior probability (hereafter the {\it prior}) that is assigned to the model's veracity based on our current understanding of the physical processes that govern the systems the model aims at explaining, and $P\left({\cal D}|{\cal M}_{i}\right)$ is the the probability of obtaining the data ${\cal D}$ given the model ${\cal M}_{i}$. The term in the denominator of Eq.~\ref{eq4} is a normalization factor (called the $\it evidence$) and is given by the summation over all possible models (i.e., summation over the different ranges of each of the model's parameters) $P\left({\cal D} \right) \equiv \Sigma_{i} P\left({\cal M}_{i}\right) P\left({\cal D}|{\cal M}_{i}\right)$. This ensures that the sum of all posterior probabilities equals unity, i.e., $\Sigma_{i} P\left({\cal M}_{i}|{\cal D}\right)=1$.
 
The specific problem we are concerned with in this work is the underlying distribution function of stellar masses in stellar clusters. If the model ${\cal M}_{i}$ can be described by a set of $N$ parameters $[\theta_{N}]_{i}$, and the data with an ensemble of stellar masses $[M_{\star}]$ of $j$ measured values $M_{\star,j}$, then the posterior probability of the model is given by:
 
\begin{equation}
P\left(\left[\theta_{N}\right]_{i}|\left[M_{\star}\right]\right)=\frac{P\left(\left[\theta_{N}\right]_{i}\right)~P\left(\left[M_{\star}\right] |\left[\theta_{N}\right]_{i}\right)}{\Sigma_{i} P\left(\left[\theta_{N}\right]_{i} \right) P\left(\left[M_{\star}\right] |\left[\theta_{N}\right]_{i})\right)}.
\label{eq5}
\end{equation}

Note that it is not necessary to calculate the evidence for comparisons of the relative probabilities of different sets of model parameters. The total likelihood of the ensemble of $N_{\star}$ measurements is given by the product of the likelihood of each measured value:

\begin{equation}
P\left(\left[M_{\star}\right] | \left[\theta_{N}\right]_{i}\right)=  \prod_{j=1}^{N_{\star}} P\left(\left[M_{\star,j}\right]  | \left[\theta_{N}\right]_{i}\right)  
\label{eq6}
\end{equation}

The prior reflects our knowledge about the physical processes that govern the values of the different parameters that describe the model. In this work we consider three distinct likelihood functions for the IMF and which are given by the tapered power-law mass function (de Marchi et al. 2004;2010; Parravano et al. 2011; Eq.~\ref{eq1}) with its 3 parameters ($\Gamma_{P}$, $M_{P}$, and $\gamma_{P}$), the lognormal+power law IMF (Chabrier 2005; Eq.~\ref{eq2}) which is described with four parameters ($\Gamma_{C}$, $M_{C}$, $\sigma_{C}$, $M_{br}$) and the three-component power law function (Kroupa 2002; Eq.~\ref{eq3}) with its five parameters ($\Gamma_{K1}$, $\Gamma_{K2}$, $\Gamma_{K3}$, $M_{K1}$, $M_{K2}$). We sample the posterior probability distribution with a Markov Chain Monte Carlo (MCMC) algorithm. The MCMC method provides a robust approach to characterize the density of samples (MCMC steps) around discrete parameters sets in the multi-dimensional parameter space. The regions with the highest density of samples correspond to regions of the parameter space with the highest probability (e.g., Gelman et al. 1996). In this work, we use an MCMC algorithm based on the Metropolis-Hastings (M-H) method (Metropolis et al. 1953; Hastings 1970). In the M-H algorithm, a random walk is performed in the parameter space from an initial, user defined position. The choice of the initial position as well as of the step size in each of the parameters can be done using an informed guess (i.e., for example in our case by looking at the binned IMFs), and/or through a "burning phase" in which the ability of the chain to converge around regions of high probability is tested. Starting from a position $[\theta_{N}](l)$ in the parameter space, a new proposed value of the parameters $[\theta_{N}](p)$ is evaluated at the next step of the chain. The new proposed position has a probability density that depends only on the current position $Q=([\theta_{N}](p)|[\theta_{N}](l))$ (we use here a symmetric uniform function such that $Q([\theta_{N}](p)|[\theta_{N}](l))=Q([\theta_{N}](l)|[\theta_{N}](p)$ ) and it is accepted, and replaced with the probability:

\begin{equation}
A([\theta_{N}](p);[\theta_{N}](l))={\rm min}\left[1,\frac{P([\theta_{N}](p)|[M_{*}])}{P([\theta_{N}](l)|[M_{*}])} \frac{Q([\theta_{N}](l)|[\theta_{N}](p)}{Q([\theta_{N}](p)|[\theta_{N}](l)}\right].
\label{eq7}
\end{equation} 

It has been shown that the ideal acceptance rate for a one-dimensional Gaussian distribution is $\approx 50 \%$ and $\approx 0.234$ for a multi-dimensional Gaussian target distribution (Gelman et al. 1997; Roberts et al. 1997). In this work, we optimize the step size in each of the parameters in order to maintain an acceptance rate of $\approx 0.234$. This should not however be understood as a very strict requirement since some of the inferred posterior distributions of the parameters are not strictly Gaussians (see below). For clusters with a small number of stars ($N_{\star}$) such as NGC 2024, we tolerate an acceptance rate of up to $40 \%$. 

\section{INTER-CLUSTER COMPARISON}\label{intercluster}

In the following, we infer the posterior probability distribution function (pPDF) of the parameters that describe the underlying IMF for the eight clusters considered in this study, and for the three likelihood functions described in \S.~\ref{models}. For each IMF model, we apply the method first to synthetic data in order to assess the ability of the method to recover the set of parameters.  

\subsection{The tapered power law model}\label{tapered}

\begin{figure*}
\begin{center}
\includegraphics[width=0.9\textwidth,height=9cm]{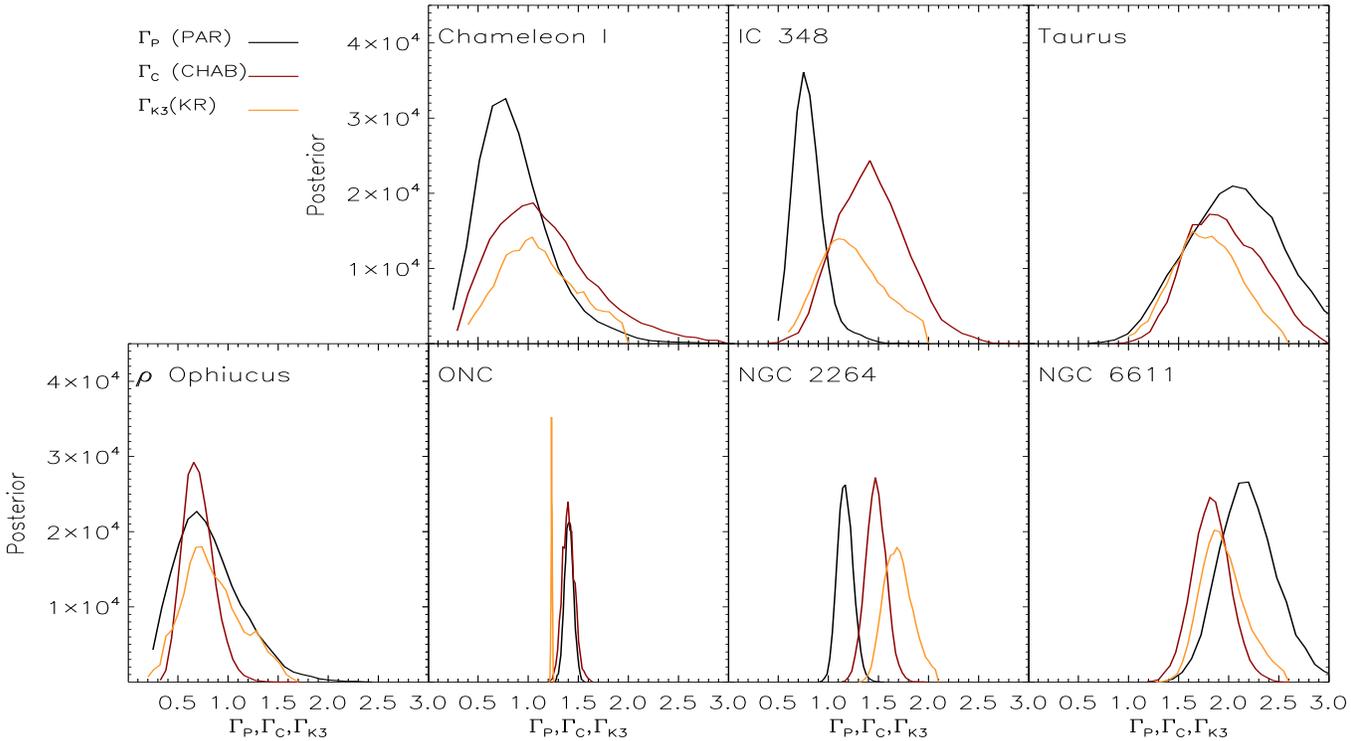}
\end{center}
\vspace{1cm}
\caption{Inter-model comparison of the pPDFs of the slope in the intermediate- to high mass between the three likelihood functions which are the tapered power law mass function (PAR; Eq.~\ref{eq1}), the lognormal+power law mass function (CHAB; Eq.~\ref{eq2}), and the three-component power law mass function (KR; Eq.~\ref{eq3}) for seven of the clusters considered in this work.}
\label{fig14}
\end{figure*}

Fig.~\ref{fig5} displays an example of the the 1-D pPDF inferred for the parameters $\Gamma_{P}$, $M_{P}$, and $\gamma_{P}$ as well as the 2-D posterior probability density in the ($\Gamma_{P}-M_{P}$), ($\Gamma_{P}-\gamma_{P}$), and ($\gamma_{P}-M_{P}$) spaces for a sample of four synthetic clusters with $N_{\star}=10^{2}$ (top left panel), $5\times 10^{2}$ (top right panel), $10^{3}$ (bottom left), and $10^{4}$ (bottom right). The $10^{2}$, $5\times10^{2}$, $10^{3}$, and $10^{4}$ stars of these clusters have been randomly sampled from the tapered power law IMF (Eq.~\ref{eq1}) in the mass range [0.02-150] M$_{\odot}$ with the Galactic field values of the parameters ($\Gamma_{P}=1.35,M_{P}=0.42,\gamma_{P}=0.57$). The dashed lines in the 1-D distributions and the white crosses in the 2-D distributions mark the position of these values in Fig.~\ref{fig5}. In the 2-D posterior probability density maps, the black, red, and orange closed contours enclose 68.27 \%, 95.45 \% and 99.73 \% (i.e., the 1-, 2- and $3\sigma$ confidence levels) of the posterior probability, respectively. In this work, we assume (both for the synthetic clusters and the real data) that we know little about the most probable value of each of the parameter. As prior probabilities on each of the parameters, we use a rectangular function given by: 

\begin{eqnarray}
\begin{array}{l} 
P([\theta_{N}])=
 \end{array}
 \left\{
 \begin{array}{l}
1;  a_{N} \leq \theta_{N} \leq b_{N} \\
\epsilon;  \theta_{N} <  a_{N}, \theta_{N} > b_{N}, \\
\end{array}
\right.  
\label{eq8}
\end{eqnarray}

where $\epsilon$ is an infinitely small quantity and where $a_{N}$ and $b_{N}$ are chosen in such a way as to bracket a large range for each of the parameters. In the inference shown in Fig.~\ref{fig5}, the values that we have used for [$a_{N},b_{N}$] are [$0.25,4$], [$0.02,1.5$], [$0.2,8$] (except for NGC 2024 where these values are [$0.25,5$], [$0.02,4$],[$-0.5,8$]), for the parameters $\Gamma_{P}$, $M_{P}$, and $\gamma_{P}$, respectively. The quantities $<\Gamma_{P}>$, $<M_{P}>$, and $<\gamma_{P}>$ that are show for each value of $N_{\star}$ are the mean values (and these are not necessarily similar to the most probable values) of the parameters and are shown along with the $1\sigma$ uncertainty which is calculated directly from the list of values of each parameter in the MCMC chain. Fig.~\ref{fig5} clearly shows that the MCMC algorithm is able to recover with a very high efficiency the values of the parameters with which the data has been generated. The figure also displays the correlations that exist between these parameters (i.e., $M_{P}$ and $\gamma_{P}$ increase and decrease with increasing values of $\Gamma_{P}$, while $M_{P}$ decreases with increasing values of $\gamma_{P}$).      

We now turn to the real clusters Taurus, Cha I, IC 348, ONC, $\rho$ Ophiuchi, NGC 2264, NGC 6611, and NGC 2024. We follow the same procedure as for the case with the synthetic data. 
In this work, the adopted concept of universality is one in which the pPDFs of each of the parameters and for all clusters will overlap at the $1\sigma$ confidence limit. This corresponds to the same confidence limit that is used for comparing the values of the parameters when fitting the binned form of the IMF. Fig.~\ref{fig6} displays the contours that enclose $68.27\%$ of the posterior probability density (i.e., at the $1\sigma$ confidence level) for each cluster. The black dashed line in the 1-D figures and the black crosses in the 2-D ones mark the position of the values of the parameters for the Milky Way field stars IMF inferred by Parravano et al. (2011). Several aspects of Fig.~\ref{fig6} are worth commenting on. As reported in studies of the IMF using binned data for Taurus and its more massive counterpart NGC 6611 (Luhman 2007; Oliveira et al. 2009), these two clusters stand out as having a steeper slope in the intermediate-to-high mass regime. The pPDF of their $\Gamma_{P}$ parameter peaks around values $> 2$ and the pPDF of their characteristic mass $M_{P}$ also peaks at larger values (0.7-0.8 M$_{\odot}$). The inferred posterior distributions of the parameters for NGC 2024 display a resemblance to those of Taurus and NGC 6611 but the inferred pPDFs for this cluster (i.e., large values of $\Gamma_{P}$ and $M_{P}$) can be understood as being the result of the absence of high mass stars in the data. However, Fig.~\ref{fig6} also shows that the cases of Taurus and NGC 6611 are not exceptions among the other clusters. Not only are the pPDFs of the inferred parameters $\Gamma_{P}$, $M_{P}$, and $\gamma_{P}$ not centered around the field IMF values, they also show little or no-overlap between these clusters at the 1$\sigma$ level (Note that the pPDF of $\Gamma_{P}$ for NGC 2024 and of $\gamma_{P}$ for the ONC are shifted by factors of -2 and -3 purely for presentation purposes). The ONC presents another case of strong deviation from the Galactic field IMF with the peak of the pPDF for the $\gamma_{P}$ parameter located at $\approx 4.5$. This striking deviation has already been noted by Da Rio et al. (2012) who has argued that the steep slope of the ONC at the low mass end cannot result from incompleteness effects in this mass regime. The 1-D pPDFs and the 2-D posterior density maps shown in Fig.~\ref{fig6} strongly suggest that there is {\it no universal set of parameters that characterizes the IMF of young clusters} and that the IMF of individual clusters can deviate, in different mass regimes, from the Galactic field IMF. The mean values of the parameters along with the $1\sigma$ uncertainties, shown in Fig.~\ref{fig6}, confirm the absence of a global overlap between the parameters of these 8 clusters at the $1\sigma$ confidence limit. The dispersion among these mean values is $0.6$ for $\Gamma_{P}$, $1.40$ for $\gamma_{P}$ ($0.25$ when excluding the ONC), and $0.27$ for $M_{P}$. Whether there is an underlying function from which these values are drawn remains open. It is however impossible to answer this question with a sample consisting of only 8 clusters, as the frequency of each of these mean values among the larger population of Galactic stellar clusters is not yet established. In order to visualize examples of the IMF constructed using accepted sets of the inferred parameters, we show in Fig.~\ref{fig7} an over-plot of 12 IMFs constructed using selected, and uncorrelated sets of the parameters ($\Gamma_{P}, M_{P}, \gamma_{P}$) from the MCMC chain to the binned IMFs for each of the clusters.             
      
\subsection{The combined lognormal-power law function}\label{lognorm}

\begin{figure*}
\begin{center}
\includegraphics[width=0.9\textwidth,height=15cm]{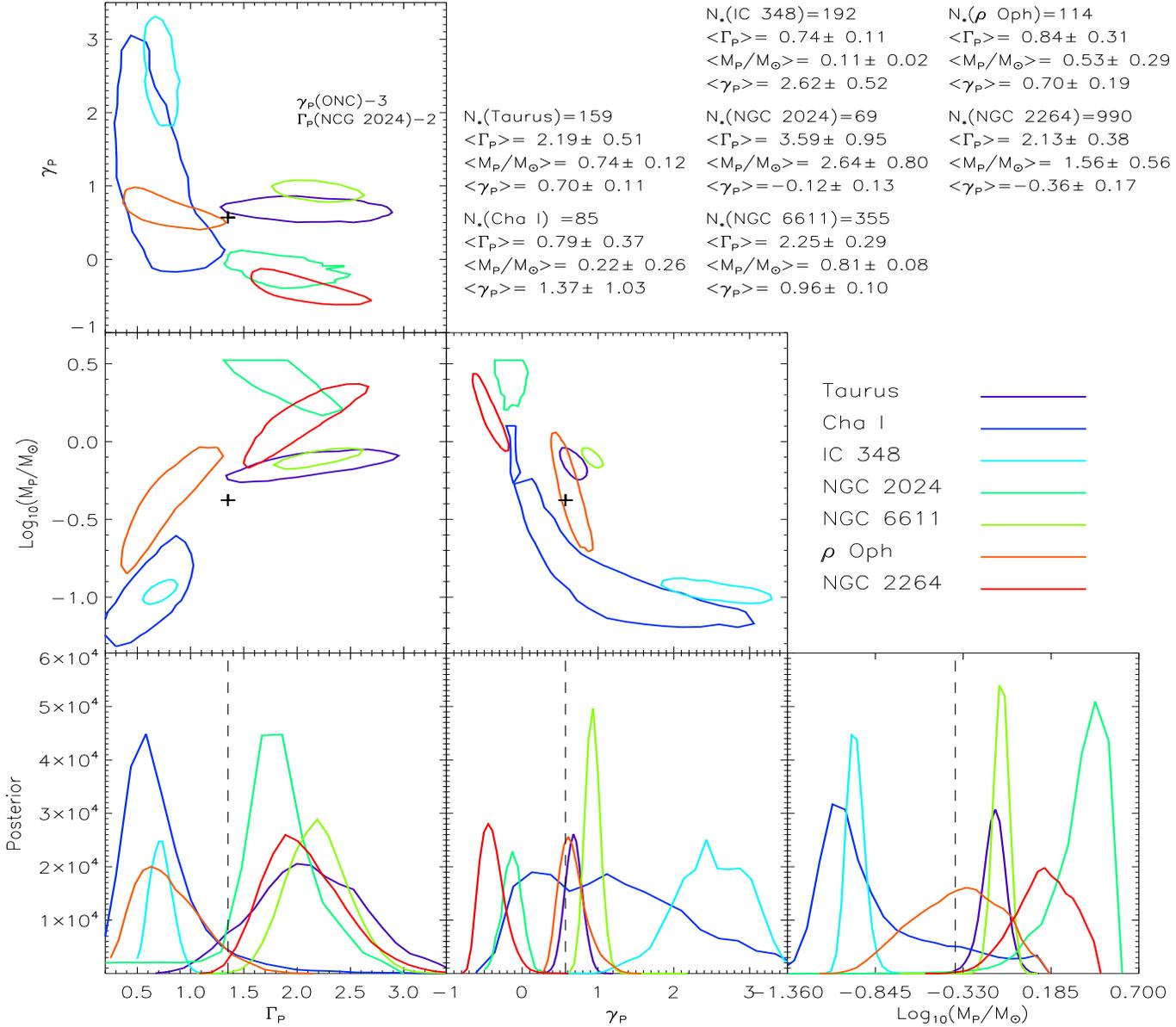}
\end{center}
\vspace{0.5cm}
\caption{Same as Fig.~\ref{fig6} but considering only stellar masses in each cluster which are higher than the completeness limit ($M_{\star} \ge M_{comp}$). The values of $M_{comp}$ of the seven clusters shown here are reported in Tab.~\ref{tab1}. The ONC is excluded from this comparison and treated separately since each star in the clusters possesses an associated completeness probability (see text in \S.~\ref{completeness} for more details).}
\label{fig15}
\end{figure*}

\begin{figure*}
\begin{center}
\includegraphics[width=0.9\textwidth,height=15cm]{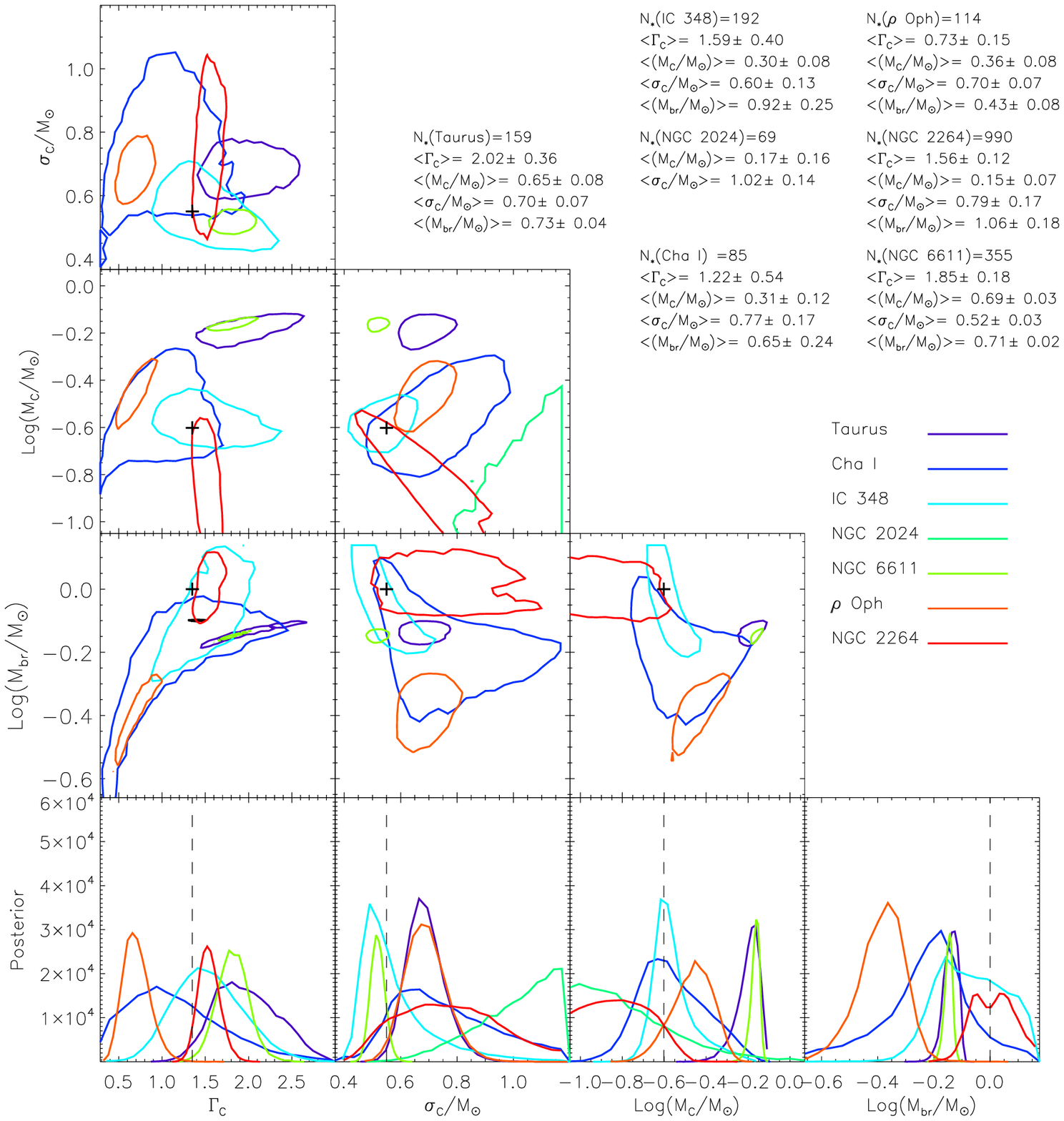}
\end{center}
\vspace{0.5cm}
\caption{Same as Fig.~\ref{fig9} but considering only stellar masses in each cluster which are higher than the completeness limit ($M_{\star} \ge M_{comp}$). The values of $M_{comp}$ of the seven clusters shown here are reported in Tab.~\ref{tab1}. The ONC is excluded from this comparison and treated separately since each star in the clusters possesses an associated completeness probability (see text in \S.~\ref{completeness} for more details).}
\label{fig16}
\end{figure*}

The inter-cluster comparison of the pPDFs of the inferred parameters that describe a tapered power law IMF clearly highlight the existence of variations among the IMF of these clusters. However, it is also relevant to explore the degree of variations among these clusters when the likelihood function has other functional forms. Here, we explore the universality of the IMF of these clusters by employing a likelihood function that is given by the combined-power law function (Chabrier 2005). The functional form of this IMF is given by Eq.~\ref{eq2} and it possesses four parameters which are the slope at the high mass end, $\Gamma_{C}$, the width of the lognormal distribution, $\sigma_{C}$, the characteristic mass, $M_{C}$, and the break point, $M_{br}$, which marks the position at which the transition occurs between the lognormal and the power-law parts of the IMF. However, only three of these parameters are independent and the value of $M_{br}$ can be calculated from any combination of $\Gamma_{C}$, $\sigma_{C}$, and $M_{C}$ (see Appendix A for details).      

As in \S.~\ref{tapered}, we start by testing the MCMC algorithm on sets of synthetic clusters. The lower right panel in Fig.~\ref{fig8} displays the IMFs in binned form of three synthetic clusters with $N_{\star}=3\times 10^{2}, 10^{3}$, and $3\times10^{3}$ and for which the individual stellar masses have been randomly sampled using the combined lognormal-power law IMF in the mass range [0.02-150] M$_{\odot}$ with the values of the parameters that correspond to the field star IMF (Chabrier 2005). An over-plot of the Chabrier (2005) IMF is also shown for comparison. Fig.~\ref{fig8} also  displays the inferred 1-D pPDFs and the 2-D posterior probability density maps for the parameters $\Gamma_{C}$, $M_{C}$, $\sigma_{C}$, and $M_{br}$ using the MCMC algorithm. For the inference of these pPDFs, we use rectangular functions (i.e., Eq.~\ref{eq8}) for the priors on $\Gamma_{C}$ and $\sigma_{C}$ with [$a_{N}-b_{N}$]=[0.25,3] and [$0.05,1.2$], respectively. We also include a prior on $M_{C}$ and $M_{br}$ by requiring that $c_{N} < M_{C}/{\rm M_{\odot}} < M_{br}/{\rm M}_{\odot} < d_{N}$, with $c_{N}=0.05$ and $d_{N}=1.5$. Overall, the recovery process of the parameters is quite robust with the peak of the pPDF of the different parameters being distributed around the values with which the data was generated within the $1\sigma$ uncertainty interval. We note however a tendency for the pPDFs to deviate from a Gaussian function and for the pPDF of $M_{br}$ to peak below the real value for clusters with the lowest numbers of stars, $N_{*}$.  

The inference of the same parameters for the eight real clusters that are considered in this work is displayed in Fig.~\ref{fig9}. As for the tapered-power law function, the colored contours associated with each of these clusters encloses $68.27\%$ of the probability in the 2-D posterior probability density maps. Similar prior functions have been used here as for the synthetic clusters. However, for the ONC, we had to reduce the ranges of the rectangular function in order to guarantee a $\sim ~23.4 \%$ acceptance rate and we required that $0.7 < \Gamma_{C} < 1.8$, $\sigma_{C}$, $ 0.05 < \sigma_{C} < 1.2$, and $ 0.1 < M_{C}/{\rm M}_{\odot} < M_{br}/{\rm M}_{\odot} < 0.8$. For NGC 2024, we only infer the pPDF of $\sigma_{C}$ and $M_{C}$ and choose rectangular prior functions for these parameters with $[a_{N},b_{N}]=[0.05,1.2]$ and [0.05,1.5], respectively. The inferred pPDFs are consistent with those obtained in \S.~\ref{tapered}. The pPDFs of $M_{C}$ and $\Gamma_{C}$ of Taurus and NGC 6611 peak at higher values than those of the other clusters, and while the pPDFs of some of the parameters for some of the clusters are broad enough to overlap with those of other clusters (e.g., the $\Gamma_{C}$ pPDFs of the Taurus and Cha I clusters), a global overlap between the pPDFs of the parameters of the clusters is not observed within the $1\sigma$ uncertainty level. The pPDFs are rather distributed around their values for the Galactic field stars IMF. At least for this sample of clusters, there is no indication that the parameters that describe their IMF are universal. As with the the case of the tapered power low function, the mean values of the parameters along with the $1\sigma$ uncertainties reported in Fig.~\ref{fig9} also confirm the absence of a global overlap among the parameters of the 8 clusters considered in this study. For a consistency check, Fig.~\ref{fig10} displays an over-plot of 12 IMFs constructed using selected and uncorrelated sets of the parameters ($\Gamma_{C}, M_{C}, \sigma_{C}$, $M_{br}$) from the MCMC chain to the binned IMF for each of the clusters.  

\subsection{The multi-component power law function}\label{mullticomp} 

\begin{figure}
\begin{center}
\includegraphics[width=\columnwidth]{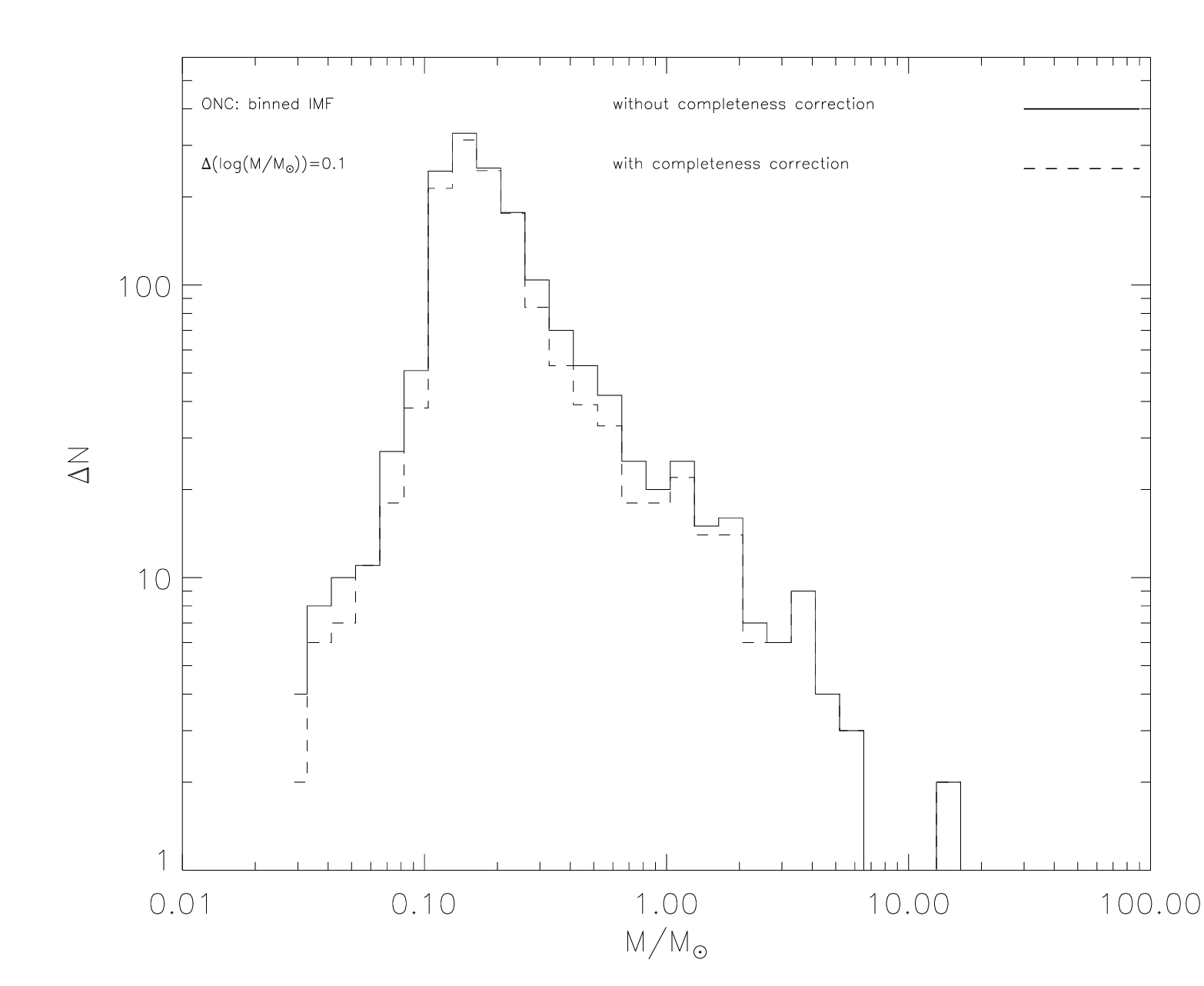}
\end{center}
\caption{Comparison of the binned form of the IMF for the ONC without and with a completeness correction. The data used here for the ONC are those derived by Da Rio et al. (2012) and which are complemented at the upper mass end by the data from Hillenbrand (1997) and Hillenbrand \& Carpenter (2000).}
\label{fig17}
\end{figure}

\begin{figure*}
\begin{center}
\includegraphics[width=0.9\textwidth,height=15cm]{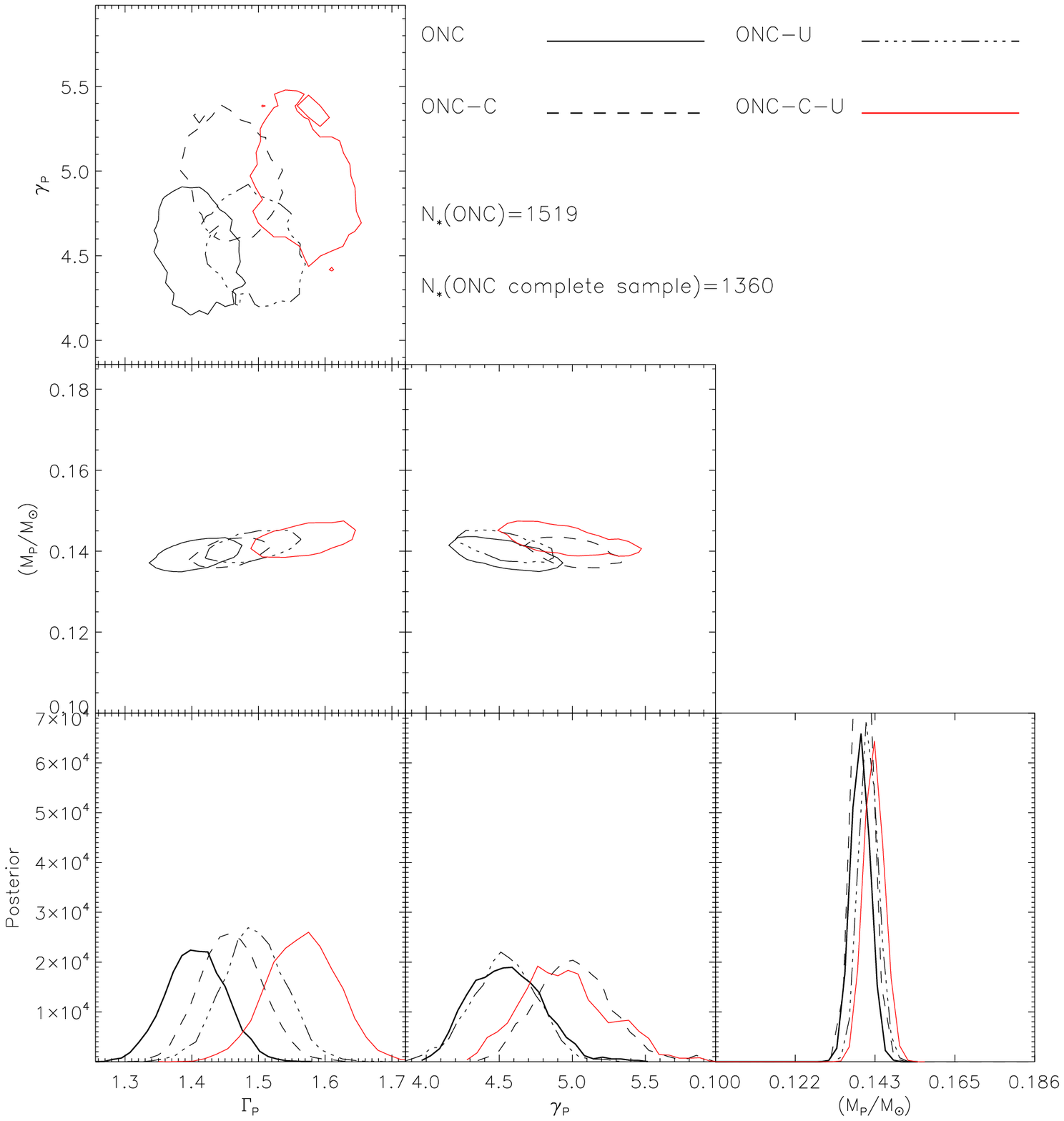}
\end{center}
\vspace{0.5cm}
\caption{The pPDFs and 2D density probability distributions of the tapered power low IMF parameters for the ONC with no corrections (ONC; full black line), with the completeness corrections (ONC-C; dashed black line), taking into account the uncertainties on individual stellar masses (ONC-U; triple dot-dashed black line) and including both a completeness correction and uncertainties on the masses (ONC-C-U; full red line).}
\label{fig18}
\end{figure*}

We complete this section by inferring the pPDFs of the parameters that describe the IMF when the likelihood function is given by the multi-component power law function IMF (i.e., Eq.~\ref{eq3}). This functional form of this IMF possesses five parameters. These are the slopes of the power laws that describe the IMF in the low-, intermediate-, and high-mass ranges,  $\Gamma_{K1}$, $\Gamma_{K2}$, and $\Gamma_{K3}$, respectively, as well as $M_{K1}$, and $M_{K2}$ which are the two break points that mark the transition from one power law to another. Only four of these parameters are independent (e.g., $\Gamma_{K1}$, $\Gamma_{K2}$, $\Gamma_{K3}$, and $M_{K1}$), and the fifth parameter ($M_{K2}$) along with the normalization can be derived using the constraints given by the total number of stars and the total mass of the cluster (see Appendix B for details). The lower right panel in Fig.~\ref{fig11} displays the binned form of the IMF for three synthetic clusters (with $N_{\star}=3\times10^{2}$, $10^{3}$, and $3\times10^{3}$) where the individual masses have been sampled from the Kroupa (2002) IMF with the fiducial Galactic field values. The Galactic field Kroupa IMF is over-plotted to the binned IMFs. Fig.~\ref{fig11} also shows the inferred 1-D pPDFs and 2-D posterior probability density maps of the parameters for these synthetic clusters. As with the other IMF models, the recovery process of the parameters is robust with the peaks of the pPDFs of the different parameters being distributed around the values with which the data was generated.

The inferred 1-D pPDFs and 2-D posterior density distributions of the parameters for the real clusters is displayed in Fig.~\ref{fig12}. The prior functions that were adopted for the different parameters are rectangular functions for $\Gamma_{K1}$, $\Gamma_{K2}$, and $\Gamma_{K3}$. The values of the step sizes in each of the parameters are chosen such that, within the limits of the adopted rectangular prior function of each of these parameters, the acceptance rate in the MCMC chains is maintained at $\approx 23.4\%$. The boundaries of the rectangular prior distribution for $\Gamma_{K1}$ were selected to fall in the range $[-1.7,0.6]$, those of $\Gamma_{K2}$ in the range $[-0.5,1]$, and those of $\Gamma_{K3}$ fall in the range $[0.2,2.6]$. We further imposed that $M_{min} < M_{K1} <e_{N}$, and $ M_{K1} < M_{K2} <  f_{N}$ where $M_{min}$ is the minimum stellar mass found in each cluster and with $e_{N}$ and $f_{N}$ falling in the range [$0.2,1$] and [$1.4,1.8$] for the different clusters, respectively. An exception to this is the case of the ONC in which we imposed $-4 <\Gamma_{K1}< -1$, $-8.5 < \Gamma_{K2} < -6$, $0.8 < \Gamma_{K3}< 2$, $M_{min} < M_{K1}<0.12$, and $M_{K1} < M_{K2}< 0.15 $. Here also, the pPDFs of $M_{K1}$ and $\Gamma_{K3}$ for Taurus and NGC 6611 are shifted to higher values with respect to the other clusters. The ONC also shows distinct pPDFs for several parameters ($\Gamma_{K1}$, $\Gamma_{K2}$, and $M_{K2}$). Note that the pPDF of $\Gamma_{K2}$ for the ONC has been shifted by +7 for visualization purposes. While the pDPFs of $\Gamma_{K1}$, $\Gamma_{K2}$, and $M_{K2}$ for the clusters (Cha I, IC 348, $\rho$ Ophiuchi, and NGC 2264) show a reasonable overlap, the pPDFs of their $\Gamma_{K3}$ are distributed around the fiducial field value of $\Gamma_{K3}=1.35$ and show a minimal level of overlap. For NGC 2024, we have chosen here to infer the entire set of parameters in order to contrast it with the its inferred pPDFs which were restricted to the low-mass in \S.~\ref{lognorm}. Fig.~\ref{fig12} shows that while the pPDFs of $\Gamma_{K1}$ and $M_{K1}$ for NGC 2024 are informative of the distribution of these parameters for this cluster, the pPDFs of of the other parameters that describe the high-mass end part are either very broad (i.e., for $M_{K2}$) or flat (i.e., for $\Gamma_{K2}$, $\Gamma_{K3}$). As for the PAR and CHAB IMF models, we show in Fig.~\ref{fig13}  an over-plot of 12 IMFs constructed using selected and uncorrelated sets of the parameters ($\Gamma_{K1}, \Gamma_{K2}, \Gamma_{K3}, M_{K1}$, $M_{K2}$) from the MCMC chain to the binned IMF for each of the clusters.  

\section{INTER-MODEL COMPARISON}\label{intermodel}

\begin{figure}
\begin{center}
\includegraphics[width=0.9\columnwidth]{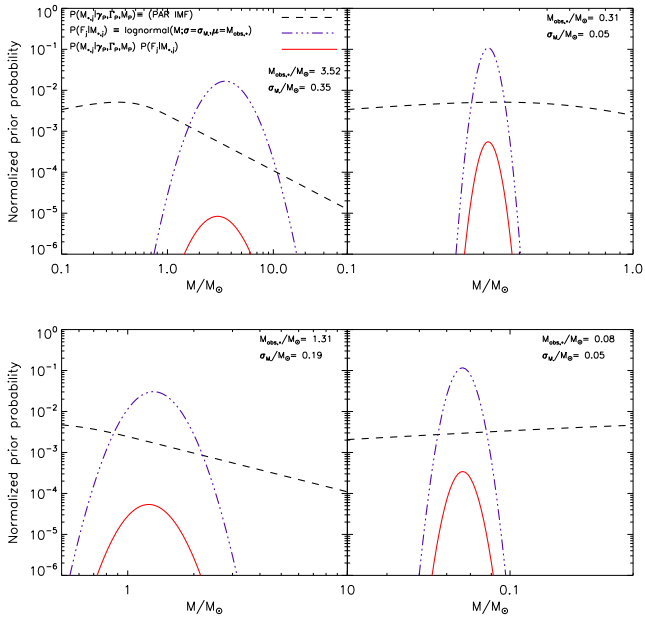}
\end{center}
\caption{Examples showing the effect of including the mass uncertainties when inferring the IMF parameters for four stars from the ONC with masses $M_{*,j}$ and uncertainty on the mass $\sigma_{M_{*,j}}$. The probability of measuring each stellar mass given its the observed flux $P(M_{*,j}|F_{*,j})$ (full red line) is given by the product of the probability of the observed flux given the mass $M_{*,j}$, $P(F_{*,j}|M_{*,j})$which is modeled by a lognormal probability function centered at $M_{*,j}$ and with a width $\sigma_{M_{*,j}}$ (blue triple dot-dashed line) with a function that describes the prior probability of measuring a particular stellar mass which we take to be the mass function itself at each iteration of the MCMC chain (black dashed line).}
\label{fig19}
\end{figure}

In this section, we make use of the information obtained above in order to compare, for each individual cluster, the pPDFs of the parameters that are common between these models. The sole parameter that is common between the three IMF models considered here is the slope in the intermediate-to-high mass end ($\Gamma_{P}$, $\Gamma_{C}$, and $\Gamma_{K3}$ for the PAR, CHAB, and KR IMF models, respectively). Fig.~\ref{fig14} displays the inferred 1-D pPDFs of $\Gamma_{P}$, $\Gamma_{C}$, and $\Gamma_{K3}$ for each of the eight clusters considered in this work. With the exception of $\rho$ Ophiuchi for which the pPDFs of $\Gamma_{P}$, $\Gamma_{C}$, and $\Gamma_{K3}$ peak at the same location, the pPDFs of $\Gamma_{P}$, $\Gamma_{C}$, and $\Gamma_{K3}$ for the remaining clusters highlight the differences between these clusters both in terms of the differences in the position of the peak value of the pPDF, their deviation from the common Galactic field value of $1.35$, as well as in terms of the width of the pPDFs of these parameters for each cluster, when their parameters are inferred using each of these models.  

\section{THE EFFECTS OF COMPLETENESS AND MASS UNCERTAINTIES}\label{completeness}

In the sections above, we have used the entire currently available stellar content of each cluster in order to infer the pPDFs of the parameters that characterize its IMF. In order to assess whether the mass functions for these clusters are truly representative of their IMF, it is important to include the effects of completeness. The completeness is in general a function of several quantities such as stellar mass, spectral energy distribution class, location in the cluster, and local effects of extinction. It is also important to include the effects of the uncertainties on the estimated masses in the Bayesian characterization of the shape of the IMF. In addition to uncertainties in the photometry, other sources of uncertainty can affect the derived masses such as uncertainties on the distance of individual stars in the cluster, extinction, and the uncertainties associated with the stellar evolutionary tracks. The effects of mass uncertainties and completeness are not expected to affect the parameters with a high level of degeneracy since they affect the stellar masses differently in different mass ranges with the effects of completeness being more significant towards the low mass end, and the mass uncertainties being more significant for higher masses. Both completeness and uncertainty on the mass are notoriously difficult to assess, and for the sample of clusters we are studying here, only the ONC have uncertainty estimates for each stellar mass measurement and an individual star-by-star completeness probability (see Da Rio et al. 2012b). The high mass stars of the ONC which are included into the ONC stellar census using the Hillenbrand (1997) data were associated a star-by-star completeness correction of 1. For the other seven clusters, there are no star-by-star completeness correction estimates, nor measurements of the uncertainties on the individual stellar masses. However, for each cluster, the authors have provided a cut-off mass above which the census of stars in each cluster is assumed to be complete. The completeness limits for each of the clusters, $M_{comp}$, is reported in Tab.~\ref{tab1}.       

We illustrate the effects of completeness and mass uncertainties using the model IMF with the smallest number of parameters which is the tapered power-law IMF model (Eq.~\ref{eq1}; De Marchi et al. 2010; Parravano et al. 2011). Except for the ONC which has a star-by-star mass uncertainty and completeness correction estimates, the other clusters considered in this work only have a cutoff mass above which the IMF is presumed to be complete (i.e., the completeness function is those case is a step function whose value is unity for stellar masses $M_{*} > M_{comp}$, and zero otherwise). A completeness correction in the form of a cutoff is obviously not an entirely satisfying approach since it results in discarding away stars with an otherwise high individual completeness probability. In Fig.~\ref{fig15}, we show the effects of completeness using this approach on the 1-D pPDFs and 2-D posterior probability density of the inferred parameters of the IMF for the seven clusters, Taurus, Cha I, IC 348, $\rho$ Ophiuchi, NGC 6611, NGC 2264, and NGC 2024. The results in Fig.~\ref{fig15} should be contrasted with those shown in Fig.~\ref{fig6} for the same clusters without any completeness correction. The loss of information from the low-mass regime results in an overall broadening of the $1\sigma$ contour levels and 1-D pPDFs. The case of Cha I is striking. The large value of $0.25$ M$_{\odot}$ for the completeness limit for the cluster NGC 2264 also results in a broadening of  the inferred 1-D pPDF of its parameters. Fig.~\ref{fig16} displays the 1-D pPDFs and 2-D posterior probability densities of the parameters when a lognormal+power law is adopted as the likelihood function. A higher degree of overlap is observed between the parameters of the 7 clusters shown here at the $1\sigma$ confidence limit (i.e., up to 5 clusters in the $M_{C}-\sigma_{C}$ parameter space). However, there is no global overlap between all clusters within the $1\sigma$ confidence interval. Our previous conclusions remain however unaffected by applying this form of completeness correction for these clusters. The 1-D pPDFs and 2-D posterior probability densities show marginal overlap and are scattered on both sides of the fiducial values of the Galactic field IMF parameters. 

We now turn to  the case of the ONC. We construct a complete sample of stars for the ONC using the star-by-star completeness probability provided by Da Rio et al. (2012b). For each star, a uniform probability value in the range [$0-1$] ($P_{[0-1,\star]}$) is drawn and compared to the star's completeness probability $P_{comp,*}$. Wherever $P_{comp,*} \geq P_{[0-1],\star}$ the star is added to the "complete" sample. The complete census of stars constructed using this approach is reduced to 1360 stars. Fig.~\ref{fig17} compares the IMF of the ONC in its binned form with and without the completeness correction. Proportionally, a larger number of low mass stars are removed from the stellar census after applying the completeness correction, leading to a narrower IMF. This is reflected in the inferred pPDFs of the slopes of the mass function at low masses, $\gamma_{P}$,  and intermediate- to high masses, $\Gamma_{P}$, which peak both at higher values (compare the full and dashed lines in Fig.~\ref{fig18}, ONC versus ONC-C). The effect of the completeness correction on the inferred pPDF of the characteristic mass is however minimal.     

In all of the results presented above, we have neglected the effect of the uncertainties associated with the individual stellar masses on the pPDFs of the inferred parameters, assuming implicitly that the measured masses are equal to their true values. In reality, the masses of stars can be affected by both observational uncertainties (e.g., instrumental uncertainties, and uncertainties on the distances of individual stars in a cluster) and uncertainties associated with the conversion of observed quantities into masses using theoretical stellar evolutionary. The observed fluxes $F_{*}$ are compared to synthetic stellar fluxes that are constructed using chosen values of several quantities such as the stellar mass ($M_{\star}$), effective temperature ($T_{eff}$), surface gravity (${\rm log}(g)$), metallicity ($Z$), age $\tau_{*}$, visual extinction ($A_{V}$), and other parameters/physical quantities that may (or not) be included in stellar evolutionary models (i.e., accretion rates in the PMS phases, rotation, etc). For a star $j$, this defines a probability for the star to have an observed flux $F_{*,j}$ given its physical quantities/parameters, $P(F_{*,j}|M_{*,j},T_{eff,j}, {\rm log}(g)_{*,j}, Z_{*,j}, A_{V,*,j}, \tau_{*,j})$. The probability of the observed flux given the mass $M_{*,j}$, $P(F_{*,j}|M_{*,j})$ can be calculated by marginalizing over the other parameters/physical quantities. The required quantity is the probability that a star has a mass $M_{*,j}$, given its observed fluxes $P(M_{*,j}|F_{*,j})$. Using the Bayes theorem, this is given by:

\begin{equation}
P(M_{*,j}|F_{*,j}) = \frac{1}{\xi} P(F_{*,j}|M_{*,j}) P(M_{*}),
\label{eq9}
\end{equation}

\noindent where $P(M_{*})$ is the prior probability function that characterizes the measurement of the stellar mass $M_{*}$, and $\xi$ is a normalization factor. As in Weisz et al. (2013), we use the stellar mass function as a prior on individual stellar masses, at each iteration in the MCMC chain. The likelihood probability of the parameters of the mass function, given the stellar mass $M_{*,j}$, is thus given by:

\begin{equation}
P(M_{*,j}|[\theta_{N}]) = \int_{M_{min}}^{M_{max}} P(F_{*,j}|M_{*,j}) P(M_{*,j}|[\theta_{N}])dM,
\label{eq10}
\end{equation}

\noindent and the likelihood probability of the parameters of the mass function, given the ensemble of stellar masses $[M_{*}]$, is thus given by:

\begin{equation}
P([M_{*}]|[\theta_{N}]) = \prod_{j=1}^{N_{\star}} \int_{M_{min}}^{M_{max}} P(F_{*,j}|M_{*,j}) P(M_{*,j}|[\theta_{N}])dM.
\label{eq11}
\end{equation}

The posterior probability distribution of each of the model parameters (in this case $\gamma_{P}$, $\Gamma_{P}$, $M_{P}$), given the stellar masses $[M_{*}]$,, is then given by:

\begin{eqnarray}
P(\gamma_{P}, \Gamma_{P}), M_{P}|[M_{*}])=\prod_{j=1}^{N_{\star}} \int_{M_{min}}^{M_{max}} P(F_{*,j}|M_{*,j}) P(M_{*,j}|[\theta_{N}])dM \nonumber \\
                             \times P(\gamma_{P}) P(\Gamma_{P}) P(M_{P}). 
\label{eq12}
\end{eqnarray}

Since the uncertainty on the mass can be due to the cumulative effects of uncertainties on the photometry, theoretical models, distances, and extinction, a reasonable assumption is to assume that the observed mass of each individual star can be described by a lognormal distribution\footnote{In case of dominant uncertainties on observables that translate strongly non-linearly in the derived stellar mass, e.g. extinction, the lognormal-shape may not be the best representation of the probability distribution. For example, if high-mass stars form behind very large column densities of gas and dust, this may cause their masses to be preferentially underestimated, in which case the probability distribution would have a broader tail at the high mass end.}. Fig.~\ref{fig18} displays four examples of the effect the existence of uncertainties in the mass has on the of the likelihood function. The modified likelihood function is given by the product of the tapered power law IMF (here shown with its fiducial Galactic field values of the parameters), and the lognormal distributions which are defined by their characteristic mass which is taken to be the observed stellar mass $M_{*,j}$ (for star $j$), and the width of the lognormal distribution is described by the measured $1\sigma$ uncertainty on the mass, $\sigma_{M_{*,j}}$. For most stars of the ONC, Da Rio et al. (2012b) provided an estimate of $\sigma_{M}$. The uncertainties reported by Da Rio et al. (2012b) on individual masses (and ages) of the stars are obtained by propagating the errors on the luminosities and effective temperatures. For the massive stars of the ONC which are not included in the Da Rio et al. census and which were taken from the Hillenbrand (1997) data, we have assigned to those stars a conservative mass uncertainty of $10 \%$ of their measured mass. This value sits at the upper end of the mass uncertainties measured by Da Rio et al. (2012). Fig.~\ref{fig18} shows that the effect of including the uncertainties on stellar masses in the inference of the IMF parameters leads to a steepening of the slope at the intermediate- to high mass end ($\Gamma_{P}$, compare ONC and ONC-U in Fig.~\ref{fig18}). This is expected since the inclusion of the IMF as a prior on measured masses tends to shift the most likely masses of each star to lower masses in this mass regime whereas the effect is negligible for masses near the peak of the IMF (i.e., see Fig.~\ref{fig19}). The effect of including the mass uncertainties has minimal effects on the shape and peak value of the pPDF of $\gamma_{P}$ and $M_{P}$. When both the completeness correction and mass uncertainties are included, the effect of shifting the peak of the pPDF of $\Gamma_{P}$ to higher masses is cumulative (i.e., red line and contours in Fig.~\ref{fig18}, case ONC-C-U). The example presented here with the ONC data highlights the importance of including the uncertainties on the mass in the inference of the pPDFs of the IMF parameters. Nonetheless, including the mass uncertainties has only significant effects on the pPDF of the slope in the intermediate- to high mass end ($\Gamma_{P}$) and minimal effects on the other two parameters ($\gamma_{P}$ and $M_{P}$). Since we observe significant variations in the pPDFs of all the IMF parameters among the clusters considered in this work (Fig.~\ref{fig6}, Fig.~\ref{fig15}, and Fig.~\ref{fig18} for the tapered power law function with and without the completeness correction), we expect that the inclusion of mass uncertainties for the remaining seven clusters would not affect our conclusions on the existence of variations between the IMF of these clusters.  Our findings are consistent with those of Weisz et al. (2013). For realistic ranges of the uncertainties on the individual stellar masses (up to $\sim 10\%$ applied across the entire considered mass range), Weisz et al. (2013) found that the peak of the pPDF for the slope at the intermediate to-high mass range is shifted by a negligible amount ($\lesssim 0.05$ dex) up to $\sim 0.1$ dex with a decreasing dynamical mass range for the considered masses. Their inferred pPDF of the slope of the  power-law shows a greater sensitivity to the completeness correction in terms of the position of the most likely value. However, they show that as long as conservative estimates of the completeness correction function are considered (i.e., a step function above which the data is assumed to be $100\%$ complete and below which the data is discarded), the inferred pPDFs of the power-low slope peaks at the same position as the one with which the mock data was produced. Our results here clearly point in the same direction as theirs. If including the completeness correction leads to an entirely complete sample, then the inferred slope of the mass function at the intermediate- to high mass regime reflects it true value.

\section{DISCUSSION}\label{discussion}

The debate over the shape of the IMF, its universality or potential variation among stellar clusters, as well as the similarity between the IMF in stellar clusters and the field stars IMF has been ongoing ever since Salpeter (1955) published his findings, and is perhaps approaching the level of myth (Melnick 2009). It is beyond the scope of this paper to review all star cluster formation models. It is important however to mention that from a purely theoretical point of view much of the arguments in favor/disfavor of variations of the IMF originate from the inclusion/absence in the models of the necessary physical processes that can lead to a significant degree of variations. A perfect illustration of this are the contrasting conclusions made by Dib et al (2010) and Hennebelle (2012). Dib et al. (2010) considered the case of  accreting protostellar cores in a non-accreting star forming clump whereas Hennebelle (2012) considered the case of non-accreting cores in an accreting clump. Dib et al. (2010) showed that the accretion of gas by protostellar cores can lead to variations in the core mass function (and hence of the IMF) when environmental conditions vary from clump-to-clump (i.e., for example when cores of the same mass have different lifetimes and accretion timescales in different environments). On the other hand, in the model of Hennebelle (2012), the accretion of gas by the clump from the larger scale environment is only expected to change the thermodynamical properties of the gas out of which newer generations of stars can form in the clump. As found by Hennebelle (2012) and earlier pointed out by Elmegreen et al. (2008), the characteristic mass of the IMF has an extremely weak dependence on the thermodynamical conditions out of which star form. The absence of accretion by the cores leaves the slopes at the high mass end and in the low mass regime unaffected.

On the observational side, Massey (2011) pointed out that much of the cluster-to-cluster variations can be due to the absence of spectroscopic measurements, the use of different reddening corrections, and different theoretical evolutionary tracks to convert luminosities into masses. He illustrated this by discussing the case of the OB association LH 58 in the LMC. For LH 58, Garmany et al. (1994) and Massey (1998) derived slopes in the high mass end of $\Gamma=1.7 \pm 0.3$ and $\Gamma=1.4 \pm 0.2$, respectively, using spectroscopy but different reddening corrections, and Hill et al. (1998) obtained $\Gamma=2.5 \pm 0.3$ using photometric measurements only, while Massey (1998) obtained $\Gamma=2$ using the Garmany et al. (1994) photometric measurements with the Hill et al. (1998) conversions from photometry to effective temperatures. At the low mass end, the case of the Taurus cluster is often cited as a potential example whose mass function deviates from the field-like IMFs (Luhman et 2000,2004a). Goodwin et al. (2004) suggested that the observed deficit of low mass stars and brown dwarfs in Taurus is due to the ejection of low mass cores by dynamical interaction while Dib et al. (2010) and Dib (2012) showed that a Taurus-like mass function can be reproduced when protostellar cores continue to accrete over longer timescales (i.e., as a result of being supported by comparatively stronger magnetic fields), and thus depleting the population of very low mass cores and shifting the peak of the mass function towards higher masses. Several works (e.g., Stolte et al. 2006; Espinoza et al. 2009, Hu$\ss$mann et al. 2012) have also reported that the slope of the IMF at the high mass end of starburst clusters such as the Arches cluster, NGC 3603, and the Quintuplet cluster might be shallower than the Salpeter value, albeit in the particular case of the Arches clusters, newer measurements have brought back the slope to a more agreeable degree with the Salpeter value (Espinoza et al. 2009; Habibi et al. 2013). Dib et al. (2007,2008) and Dib (2007) proposed that shallower-than Salpeter slopes can result from the efficient coalescence of closely packed protostellar cores in a dense protocluster environment. Hocuk \& Spaans (2011) have also reported that the characteristic mass of the IMF can be shifted to higher values and that a shallower than-Salpeter slope in the intermediate- to high mass regimes can be reproduced in environments where the gas is exposed to very high X-ray and cosmic ray fluxes.   

\section{CONCLUSIONS}\label{conclusions}

In this work, we have presented the first analysis of the universality of the stellar initial mass function (IMF) using Bayesian statistics with a Monte Carlo Markov Chain method for a sample of eight young Galactic stellar clusters (IC 348, ONC, NGC 2024, NGC 6611, NGC 2264, $\rho$ Ophiuchi, Chameleon I, and Taurus). Those eight clusters were chosen based on the availability of their measured individual stellar masses. With the exception of the ONC, individual masses where derived using the same theoretical evolutionary tracks, at least in the low mass end regime. We derive the posterior probability distributions (pPDFs) of the parameters that describe the IMF all the way from the brown dwarfs and low mass regime up to the regime of massive stars. For each of the considered clusters, we infer the pPDF of the IMF parameters when the likelihood function that describes the IMF is given by a tapered power law function (De Marchi et al. 2010; Parravano et al. 2012), a lognormal distribution at low masses coupled to a power law at higher masses (Chabrier 2005), and a multi-component power law function (Kroupa 2002). The tapered power function has three parameters which are the slopes of the mass function at the low and intermediate- to high mass regimes ($\gamma_{P}$ and $\Gamma_{P}$, respectively) and the characteristic mass $M_{P}$. The lognormal-power low function is described by four parameters which are the width of the lognormal distribution and its characteristic mass ($\sigma_{C}$ and $M_{C}$, respectively), the slope of the power law component, $\Gamma_{C}$, and the position of the break point between the lognormal and power law parts of the mass function, $M_{br}$. The three-component mass function has five parameters which are the three slopes of the power law components ($\Gamma_{K1}, \Gamma_{K2}$, $\Gamma_{K3}$) and the two break points that mark the transition from one power law to another.

The inter-cluster comparison of the inferred 1D pPDFs as well as the 2D surface density probabilities of the parameters with the different likelihood functions ($[\gamma_{P}, \Gamma_{P}, M_{P}$]; [$\sigma_{C}, M_{C}, \Gamma_{C}, M_{br}$], [$\Gamma_{K1}, \Gamma_{K2}, \Gamma_{K3}, M_{K1}, M_{K2}$]) shows that these distributions do not overlap within the $1\sigma$ uncertainty level (Fig.~\ref{fig6}, Fig.~\ref{fig9}, and Fig.~\ref{fig12}). When the likelihood function is the tapered power law mass function, the most likely values of $\gamma_{P}$ span the range $\approx$ $-0.3$ to $1.6$ with NGC 2024 and the ONC possessing these two extreme values, respectively. The pPDFs of $\Gamma_{P}$ also show variations over a significant range going from $\approx 0.6-0.8$ for $\rho$ Oph, IC 348, and Cha I, and up to $\approx 2.2-2.3$ for Taurus and its most massive counterpart NGC 6611. Variations in the most likely value of the inferred pPDFs for $M_{P}$ also span a significant range from $\approx 0.12$ M$_{\odot}$ for the ONC and IC 348 and up to $0.71-0.8$ M$_{\odot}$ for Taurus and NGC 6611. The inferred pPDF of $M_{P}$ which peaks at a much higher value of $\approx 2.2$ M$_{\odot}$ should not be considered very seriously since the current census of stars for NGC 2024 and which we used in this work contains only low mass stars with a maximum mass of $\approx 0.72$ M$_{\odot}$. Similar variations are observed when the likelihood functions that describes the IMF is chosen to be the lognormal+power law function (Fig.~\ref{fig9}) or the multi-component power law function (Fig.~\ref{fig12}). Furthermore, the most probable values of the IMF parameters for most of these clusters deviate substantially from their values for the Galactic field stars IMF. 

We have explored the effects of mass incompleteness on the inferred pPDFs of the IMF parameters. The completeness correction whether performed using a star-to-star completeness probability (available only for the ONC), or by imposing a cutoff mass at the low mass end above which the mass function is assumed to be complete (i.e., for the remaining seven clusters) affects essentially the inferred pPDFs of the slope of the mass function at the low mass end (i.e., compare the pPDFs of $\gamma_{P}$ in Fig.~\ref{fig6} with those in Fig.~\ref{fig15} and Fig.~\ref{fig18}). We have also measured the effect on the inferred pPDFs of taking into account the uncertainties on the measured masses which we model as lognormal distributions. We also chose the IMF itself (at each iteration in the MCMC chain) as a prior on the measured masses. This is performed here only for the case of the ONC for which such uncertainties are available. Our results (Fig.~\ref{fig18}) show that the inclusion of the mass uncertainties shifts the pPDF of the slope at the intermediate- to high mass end $\Gamma_{P}$ to higher masses, leaving the pPDFs of $\gamma_{P}$ and $M_{P}$ unaffected. This is expected due to the combined effect that uncertainties on the stellar masses are higher for higher mass stars and the fact that the mass function has a negative slope in the high mass regime. 

The results presented in this paper highlight the importance of using advanced and robust statistical methods for the inference of the parameters that characterize the stellar initial mass function. We find that all of the three IMF functional forms are able to reproduce the observations if the parameters that define each function are allowed to vary significantly (i.e., beyond the $1\sigma$ limit). However, it is not clear which one of these three functions, if any, captures the physics of the star formation process. In this work, we have inferred the parameters that strictly describe the "shape" of the IMF. Additional parameters can be included such as the mass of the cluster (see e.g., Cervi\~{n}o et al. 2013) maximum stellar mass present in the cluster in the framework of a cluster mass-maximum stellar mass relation. Since we are dealing here with only eight clusters with a small range in cluster masses, the significance of including $M_{max}$ as a free parameter in this study is rather limited. In this work, we have used rather simple prior functions on the parameters (i.e., rectangular functions), thus assuming implicitly that we have little knowledge on their parameters from theory and or theoretical models of star formation. At least for the eight clusters considered in this work, the posterior probability distribution functions derived for their parameters can serve as prior functions when using improved data from future surveys. The high sensitivity and angular resolution of the {\it James Web Space Telescope} (JWST) will open entirely new and exciting perspective for the application of this method to a large set of Galactic and extragalactic clusters with resolved stellar populations.    

\section*{Acknowledgments}

I thank the Referee for useful comments and suggestions. I enjoyed discussions on Bayesian statistics with Daniel Mortlock and Kaisey Mandel. I am very grateful to Kevin Luhman, Hwankyung Sung, Catarina Alves de Oliveira, Nicola Da Rio, and Joana Oliveira for sharing their data and for useful information. I also thank Sacha Hony for a careful reading of the manuscript and useful suggestions. This work was supported by a Marie-Curie Intra European Fellowship under the European Community's Seventh Framework Program FP7/2007-2013 grant agreement no 627008. 
 
{}

\appendix 

\section{DERIVING THE NON-INDEPENDANT VARIABLES IN THE COMPOSITE LOGNORMAL+POWER LAW MASS FUNCTION }

In this appendix, we derive the value of the break point $M_{br}$ and normalisation $A_{C}$ in the composite lognormal+power law IMF when the other parameters, $M_{C}$, $\Gamma_{C}$, and $\sigma_{C}$ are given. The value of $M_{br}$ is given by the equality:

\begin{equation}
0.041\times M_{br}^{-\Gamma_{C}} = 0.076\times{\rm exp}\left\{ -\frac{({\rm log}M_{br} - {\rm log}M_{c})^{2}}{2 \sigma_{c}^{2}} \right \}.
\label{eqa1}
\end{equation}

After a few developments, one can show that this leads to a quadratic equation in ${\rm log}M_{br}$:

\begin{equation}
({\rm log}M_{br})^{2}-{\rm log}M_{br} \left(\frac{2 \sigma_{C}^{2} \Gamma_{C} }{{\rm log}e}+2{\rm log}M_{c}\right)+\left(({\rm log}M_{C})^{2}- 2 \sigma_{C}^{2}\frac{{\rm log}1.85}{{\rm log}e}\right).
\label{eqa2}
\end{equation} 

The discriminant of Eq.~\ref{eqa2} is given by:

\begin{equation}
 \Delta=\frac{4 \sigma_{C}^{2}}{{\rm log}e} \left(\sigma_{C}^{2} \Gamma_{C}^{2}+ 2 \Gamma_{C} {\rm log}M_{C} {\rm log}e+2~{\rm log}1.85~{\rm log}e\right). 
\label{eqa3}
\end{equation}

In the MCMC chain, we retain steps that satisfy ${\sqrt{\Delta}} > 0$. For each step at which this condition is satisfied, Eq.~\ref{eqa2} has two solutions which are given by:  

\begin{equation}
{\rm log}M_{br,1}=\frac{\sigma_{C}^{2}\Gamma_{C}}{{\rm log}e}+{\rm log}M_{C}+\sqrt{\Delta},
\label{eqa4}
\end{equation} 

\noindent and,

\begin{equation}
{\rm log}M_{br,2}=\frac{\sigma_{C}^{2}\Gamma_{C}}{{\rm log}e}+{\rm log}M_{C}-\sqrt{\Delta}.
\label{eqa5}
\end{equation} 

\noindent We further require that $M_{br} > M_{c}$ and in the MCMC chain we assign to $M_{br}$ the value of $min \left(M_{br1},M_{br,2}\right)$ that satisfies this condition. We find that for all parameters that satisfy $\sqrt{\Delta} > 0$, $min \left(M_{br1,M_{br,2}}\right)=M_{br,2}$. With knowledge of $M_{br}$, $\sigma_{C}$, $\Gamma_{C}$, and $M_{C}$, it is straightforward to derive the normalisation term $A_{c}$ for each permutation of these parameters in the Monte Carlo chain using the total mass of the cluster. 
 
\section{DERIVING THE NON-INDEPENDANT VARIABLES IN THE THREE-COMPONENT MASS FUNCTION } 

In this appendix, we derive the values of the non-independent parameters in the three-component power law mass function, when the values of the independent variables $\Gamma_{K1}$, $\Gamma_{K2}$, and $\Gamma_{K3}$, and $M_{K2}$ vary in the Monte Carlo chain. For each set of values of [$\Gamma_{K1}$,$\Gamma_{K2}$,$\Gamma_{K3}$, M$_{K1}$], the values of $M_{K2}$, and $A_{K}$ can be derived using the following set of equations:

\begin{eqnarray}
N_{*} & = & \int_{M_{min}}^{M_{max}} \psi_{K} (M) dM = A_{K} \int_{M_{min}}^{M_{K1}} \left(\frac{M}{M_{K1}}\right)^{-\alpha_{K1}} dM \nonumber \\ 
           & + &  A_{K} \int_{M_{K1}}^{M_{K2}} \left(\frac{M}{M_{K1}}\right)^{-\alpha_{K2}} dM  \nonumber \\ 
           & + &  A_{K} \int_{M_{K2}}^{M_{max}} \left(\frac{M_{K2}}{M_{K1}}\right)^{-\alpha_{K2}} \left(\frac{M}{M_{K2}}\right)^{-\alpha_{K3}} dM,
\label{eqb1}
\end{eqnarray}

\begin{eqnarray}
M_{cl} & = & \int_{M_{min}}^{M_{max}} M \psi_{K} (M) dM = A_{K} \int_{M_{min}}^{M_{K1}} M \left(\frac{M}{M_{K1}}\right)^{-\alpha_{K1}} dM \nonumber \\ 
           &  + & A_{K} \int_{M_{K1}}^{M_{K2}} M \left(\frac{M}{M_{K1}}\right)^{-\alpha_{K2}} dM \nonumber \\ 
           &  + & A_{K} \int_{M_{K2}}^{M_{max}} M \left(\frac{M_{K2}}{M_{K1}}\right)^{-\alpha_{K2}} \left(\frac{M}{M_{K2}}\right)^{-\alpha_{K3}} dM,
\label{eqb2}
\end{eqnarray}

\noindent where $\alpha_{K1}=\Gamma_{K1}+1$, $\alpha_{K2}=\Gamma_{K2}+1$, $\alpha_{K3}=\Gamma_{K3}+1$, $N_{*}$ is the number of stars in the cluster, $M_{cl}$ is the mass of the cluster, and $M_{min}$ and $M_{max}$ are the minimum and maximum masses of stars present in the cluster, respectively. Developing these integrals and replacing $(\alpha_{K1},\alpha_{K2},\alpha_{K3})$ by $(\Gamma_{K1},\Gamma_{K2},\Gamma_{K3})$ yields to the following set of equations:

\begin{eqnarray}
 A_{K} \left\{  \frac{1}{M_{K1}^{-\Gamma_{K1}-1}} \frac{1}{\Gamma_{K1}} \left[M_{min}^{-\Gamma_{K1}}-M_{K1}^{-\Gamma_{K1}}\right]  \right. &  \nonumber  \\  
          +\frac{1}{M_{K1}^{-\Gamma_{K2}-1}} \frac{1}{\Gamma_{K2}} \left[M_{K1}^{-\Gamma_{K2}}-M_{K2}^{-\Gamma_{K2}}\right]                         &  \nonumber \\
          +\frac{1}{M_{K2}^{-\Gamma_{K3}-1}} \frac{1}{\Gamma_{K3}} \left[M_{K2}^{-\Gamma_{K3}}-M_{max}^{-\Gamma_{K3}}\right] &\left. \left(\frac{M_{K2}}{M_{K1}}\right)^{-\Gamma_{K2}-1} \right\} \nonumber \\ 
          -N_{*}=0 & 
\label{eqb3}
\end{eqnarray}
 
\begin{eqnarray}
A_{K} \left\{  \frac{1}{M_{K1}^{-\Gamma_{K1}-1}}  \frac{1}{1-\Gamma_{K1}} \left[M_{K1}^{1-\Gamma_{K1}}-M_{min}^{1-\Gamma_{K1}} \right] \right.    &   \nonumber \\
        +\frac{1}{M_{K1}^{-\Gamma_{K2}-1}} \frac{1}{1-\Gamma_{K2}} \left[M_{K2}^{1-\Gamma_{K2}}-M_{K1}^{1-\Gamma_{K2}} \right]                            &   \nonumber \\  
        +\frac{1} {M_{K2}^{-\Gamma_{K3}-1}} \frac{1}{1-\Gamma_{K3}} \left[M_{max}^{1-\Gamma_{K3}}-M_{K2}^{1-\Gamma_{K3}} \right]&\left.\left( \frac{M_{K2}}{M_{K1}}\right)^{-\Gamma_{K2}-1}\right\} \nonumber \\ 
        -M_{cl}= 0, &  
\label{eqb4}
\end{eqnarray}

With $\Gamma_{K1}$, $\Gamma_{K2}$, $\Gamma_{K3}$, $M_{K1}$ known at each step in the chain and with $M_{min}$, $M_{max}$, $M_{cl}$, and $N_{*}$ constant for each cluster, we solve Eqs.~\ref{eqb3}  and \ref{eqb4} using a modified version of Powell algorithm for functions minimisation (Powell 1964; Press et al. 2007) and derive the values of $M_{K2}$ and $A_{K}$. 
 
\label{lastpage}

\end{document}